\documentclass[12pt,preprint]{aastex}
\usepackage{amssymb, amsmath, fullpage, graphicx,bm}
\shorttitle{RCB stars from double degenerate white dwarf mergers}
\shortauthors{Staff et al.}

\begin{document}
\title{The role of dredge-up in double white dwarf mergers}
\author{Jan. E. Staff}
\affil{College of Science and Math, University of the Virgin Islands, St
Thomas, 00802, United States Virgin Islands}
\author{Brandon Wiggins}
\affil{Center for Theoretical Astrophysics, Los Alamos National Laboratory, Los Alamos, NM 87545, USA and\\
Southern Utah University, 351 W University Blvd, Cedar City, UT
84720, USA}
\author{Dominic Marcello}
\affil{Department of Physics and Astronomy, Louisiana State University,
202 Nicholson Hall, Tower Dr., Baton Rouge, LA 70803-4001, 
USA}
\author{Patrick M. Motl}
\affil{School of Sciences,
Indiana University Kokomo, Kokomo, Indiana 46904-9003, USA}
\author{Wesley Even and Chris L. Fryer}
\affil{Center for Theoretical Astrophysics, Los Alamos National Laboratory, Los Alamos, NM 87545, USA}
\author{Cody Raskin}
\affil{Lawrence Livermore National Laboratory, P.O. Box 808, L-038,
Livermore, CA 94550, USA}
\author{Geoffrey C. Clayton and Juhan Frank}
\affil{Department of Physics and Astronomy, Louisiana State University,
202 Nicholson Hall, Tower Dr., Baton Rouge, LA 70803-4001,
USA}

\date{}

%
%

\begin{abstract} 

We present the results of an investigation of the dredge-up and mixing
during the merger of two white dwarfs with different chemical compositions
by conducting hydrodynamic simulations of binary mergers for
three representative mass ratios.  
In all the simulations, the total mass of the two white
dwarfs is  $\lesssim1.0~{\rm M_\odot}$.  Mergers involving a CO and a He
white dwarf have been suggested as a possible formation channel for R Coronae
Borealis type stars, and we are interested in testing if such mergers 
lead to conditions and outcomes in agreement with observations.  
Even if the conditions during the merger and subsequent nucleosynthesis
favor the production of  $^{18}{\mathrm O}$, the merger must avoid dredging 
up large amounts of $^{16}{\mathrm O}$, or else it will be difficult to 
produce sufficient $^{18}{\mathrm O}$
to explain the oxygen ratio observed to be of order unity.  
We performed a total of 9 simulations using
two different grid-based hydrodynamics codes using fixed and adaptive meshes,
and one smooth particle
hydrodynamics (SPH) code. We find that in most of the simulations, $>10^{-2}~{\rm M_\odot}$ of $^{16}{\mathrm O}$ is indeed dredged
up during the merger. However, in SPH simulations where the accretor is a 
hybrid He/CO white dwarf
with a $\sim 0.1~{\rm M_\odot}$ layer of helium on top, we find that 
no $^{16}{\mathrm O}$ is being dredged up, while in the $q=0.8$ simulation $<10^{-4}~{\rm M_\odot}$ of $^{16}{\mathrm O}$ has been brought up, making a WD binary consisting of a hybrid CO/He WD and a companion He WD an excellent candidate for the progenitor of RCB stars.

\end{abstract}

%
%
\section{Introduction} 

R Coronae Borealis (RCB) stars are hydrogen deficient, supergiant
stars, with distinguishing chemical abundance patterns and photometric properties.  
They are carbon rich, observed to consist mainly of helium with
$1\%$ carbon by mass \citep{clayton96,clayton12}.  They produce clumps of
dust that may obscure our view \citep{okeefe39}, making them fade at
irregular intervals by up to 8 magnitudes over a short timescale (weeks),
and then gradually re-brighten on a timescale of months to years.  Cool RCB
stars have not been observed in binary systems, making it difficult to
measure the mass.  Based on stellar pulsation models, the mass is assumed to
be of the order $1~{\rm M_\odot}$ \citep{saio08,han98}.

The oxygen isotopic ratio $^{16}\mathrm{O}$ to $^{18}\mathrm{O}$ has been
found to be of the order unity in RCB stars \citep[measured to be between
0.3 and 20:][]{clayton07,garcia09,garcia10}.  In the solar neighborhood,
this ratio is found to be $\sim500$ \citep{scott06}, which is also a typical
value in the Galactic interstellar medium \citep{wilson94}.  What causes
this unusually low oxygen ratio in RCB stars?

In a single star, partial He burning can produce $^{18}\mathrm{O}$, but if
the process continues $^{18}\mathrm{O}$ will be turned into
$^{22}\mathrm{Ne}$ \citep{clayton05}.  Likewise, proton capture processes: 
$^{17}{\mathrm O}({\rm p, \gamma})^{18}{\mathrm F}({\rm
\beta^+})^{18}{\mathrm O}$ can also produce $^{18}{\mathrm O}$, but this
process also continues ($^{18}{\mathrm O}({\rm p, \gamma})^{19}{\mathrm
F}$). Two scenarios for the progenitors of
RCB stars are often discussed; the final helium shell flash and the double
degenerate white dwarf (WD) merger \citep{webbink84,renzini90}.
A variation of the double degenerate merger model is that the merger is
between two He WDs. However, \citet{zhang12} found that less than $1\%$ of
RCB stars may be formed this way.
No overproduction of $^{18}\mathrm{O}$ is expected in the final-flash
scenario but in a WD merger, partial helium burning may take place leading
to enhanced $^{18}\mathrm{O}$.

In the double degenerate merger scenario, two situations are possible: a
cold merger where no nucleosynthesis occurs, or a hot merger where
additional nucleosynthesis takes place.  Using non-LTE model atmospheres,
\citet{pandey11} argued that a cold merger (ie.  no additional
nucleosynthesis) could explain the observed oxygen abundances.  Based on one
dimensional stellar evolution models, \citet{jeffery11} found that both cold
and hot mergers could result in the oxygen surface abundances that can
explain the observations.  In \citet{staff12} we presented the results of
several grid-based hydrodynamics simulations of the merger of a 
He-WD (the donor) and a
CO-WD (the accretor) with varying mass ratio but constant total mass of $\sim0.9~{\rm M_\odot}$.  We found that for mass ratio
$q>0.7$ ($q={\rm M_{donor}}/{\rm M_{accretor}}$), the temperature does not get sufficiently high to allow much
nucleosynthesis to happen.  However, for $q\lesssim0.7$, we found that a hot
``shell of fire'' (SOF) formed, with temperatures up to and above
$2\times10^8~{\rm K}$, with lower mass ratio leading to higher temperatures. 
The densities in the SOF are around $10^5~{\rm g~cm^{-3}}$.  Under these
conditions, helium can start burning.  It is therefore in the SOF that
$^{18}{\mathrm O}$ can form, which we found is needed in order to get the
unusual oxygen ratios seen in RCB stars. \citet{zhang14} investigated the
post-merger evolution of CO+He WD binaries, and found that to explain the
$^{12}\mathrm{C}$ abundances in RCB stars, the accretor mass must be around
$0.55~{\rm M_\odot}$, while the He WD must have a mass $>0.3~{\rm M_\odot}$. 
They also found that $^{18}\mathrm{O}$ can be produced and can survive, to
account for the observed oxygen ratio.

\citet{staff12} found that under very special conditions, it was possible
to achieve an oxygen ratio of $\sim4$ (comparable to observed oxygen
ratios), however these conditions may be difficult to achieve in nature (they found this in their $q=0.7$ simulation, the same as the non-AMR $q=0.7$ simulation presented here, by assuming that density, temperature, etc. remains unchanged in the SOF for $\sim 100$ years). 
\citet{longland11}, using SPH simulations of WD mergers, also found it
difficult to achieve the unusually low oxygen ratio found in RCB stars. 
Using an idealized post-merger configuration based on the hydrodynamic
simulations of WD mergers in \citet{staff12}, \citet{menon13} were able to
reproduce the observed abundance ratios in RCB stars using a one-dimensional
stellar evolution code. In \citet{menon13} a four-zone model was assumed, consisting of the cold core of the merged object, the SOF, a thin buffer zone between them, and the relatively cold envelope. They assumed that most of the dredged up material from the accretor ended up in the buffer zone. Therefore they could find lower $^{16}\mathrm{O}$ to $^{18}\mathrm{O}$ ratios than that of \citet{staff12}, in better agreement with observations.

The issue is not that $^{18}\mathrm{O}$
is not produced, but rather that much $^{16}\mathrm{O}$ is being dredged up
from the accretor (the oxygen in a CO WD is mainly $^{16}\mathrm{O}$) into
the SOF where the $^{18}\mathrm{O}$ can form. 
Hence, in order to get an oxygen ratio of order unity, as much
$^{18}\mathrm{O}$ must be produced as $^{16}\mathrm{O}$ is being dredged up.
Therefore we are interested in particular in how much $^{16}{\mathrm O}$ is 
being dredged up from the accretor to the SOF. As the $^{18}{\mathrm O}$
must be formed in the SOF, it must be brought from the SOF up into the
atmosphere of the star where it is observed. However, it is implausible that
only $^{18}{\mathrm O}$ is brought up from the SOF, and not 
$^{16}{\mathrm O}$ also present at the same place. Hence the
oxygen ratio in the SOF will to some extent reflect the observed oxygen
ratio\footnote{The observed $^{16}{\mathrm O}$ to $^{18}{\mathrm O}$ ratio
may be higher than in the SOF, since there may be some $^{16}{\mathrm O}$
present in the atmosphere prior to the production of $^{18}{\mathrm O}$ in the SOF, that can further dilute the ratio.}.

In this paper, we present the results of simulations of the merger of two WDs.
We have simulated the merger of three different mass ratios: $q=0.5$, $q=0.7$, and $q=0.8$ with total mass $M_{\rm tot}<1~{\rm M_\odot}$, using three different simulation codes.
Many binary WD systems are known, and also systems that will merge within a Hubble time. \citet{kilic11} lists 12 such binary systems where at least one component is a He WD with $M<0.25~{\rm M_\odot}$. 
He WD must form through common envelope interactions (with an envelope ejection), and hence must be in short period binaries, as the main sequence lifetime for low mass stars that would lead to He WDs are much longer than a Hubble time. 
Many He WDs are in binary systems with another WD. SDSS J092345.60+302805.0 \citep{brown10} has one WD with mass of $0.23~{\rm M_\odot}$ and the most likely mass of the other (with an inclination of $60^\circ$) is $0.44~{\rm M_\odot}$, which will merge in $\sim 130\times10^6~{\rm yrs}$. 
This is very similar to our simulated system with $q=0.5$ in this paper. 
J1436+5010 \citep{mullally09} has one component with mass $M=0.24~{\rm M_\odot}$, and the other has a mass $>0.46~{\rm M_\odot}$ (but for $i=60^\circ$ the other component has mass $M=0.60~{\rm M_\odot}$) that will merge in $<100\times10^6~{\rm yrs}$, so this system may also be similar to our $q=0.5$ simulated system.
\citet{hermes12} reported on SDSS J065133.338+284423.37, a binary WD system with masses of $0.26~{\rm M_\odot}$ and $0.50~{\rm M_\odot}$ and an orbital period of 12.75 minutes, which is also close to the masses in our $q=0.5$ simulation.
\citet{nelemans05} observed five close WD binary systems. 
One of these, WD1013-010, has one component with mass $0.44~{\rm M_\odot}$, while the component has a mass $>0.38~{\rm M_\odot}$ with an orbital period of $0.44~{\rm days}$. 
These masses are not too dissimilar from our simulated $q=0.8$ system.
Another binary WD system, SDSS J104336.275+055149.90, reported by \citet{brown17}, has WD with masses of $0.30~{\rm M_\odot}$ and $0.52~{\rm M_\odot}$, and is expected to merge in $20\times10^6~{\rm yrs}$.
The masses in this system is not that dissimilar from our simulated $q=0.7$ system.

In this paper we want to investigate the dredge-up of $^{16}\mathrm{O}$ from
the core of the accretor into the SOF in more detail.  We will investigate
how much $^{16}{\mathrm O}$ is at densities below $\rho<10^{5.2}~{\rm
g~cm^{-3}}$, and at $\rho<10^{5}~{\rm g~cm^{-3}}$ for each of the
simulations in this paper.  
These fiducial values are chosen to represent approximately the transition
 between the core and the envelope of the accretor.
 The numerical tools used in this paper are
described in section~\ref{methodsection}.
We will compare the results of three different
hydrodynamics codes, one fixed-grid-based code (See \S 2.1),  one 
grid-based code with Adaptive Mesh Refinement (AMR) (See \S 2.2), 
and one SPH code (See \S 2.3) to test if the numerical method affects the results.   
 We investigate three
different mass ratios, $q=0.5$ with $M_{\rm tot}=0.7~{\rm M_\odot}$, $q=0.7$
with $M_{\rm tot}=0.9~{\rm M_\odot}$ \citep[similar to one of the
simulations in][]{staff12}, and $q=0.8$ with $M_{\rm tot}=0.9~{\rm
M_\odot}$.  The accretor in the $q=0.5$ and $q=0.8$ are ``hybrid'' WDs, that
is they have a CO core and a thick layer of $0.1~{\rm M_\odot}$ He on top. 
Perhaps this layer of He can prevent dredge-up of $^{16}\mathrm{O}$ from the
core?  Our results are presented in section~\ref{resultssection}, we have a
discussion of our results and of other recent work on similar topics in section~\ref{discussionsection}, and finally we conclude in
section~\ref{conclusionsection}.

%
%

\section{Methods}
\label{methodsection}

Using three different codes, we investigate and compare three different sets
of initial conditions: $q=0.7$ and $M_{\rm tot}=0.9~{\rm M_\odot}$, $q=0.5$
and $M_{\rm tot}=0.71~{\rm M_\odot}$ where the accretor is a hybrid WD of
which $0.13~{\rm M_\odot}$ is $^4{\rm He}$, and
$q=0.8$ and $M_{\rm tot}=0.9~{\rm M_\odot}$ where the accretor is a slightly
more massive hybrid WD of mass $0.5~{\rm M_\odot}$, of which $0.13~{\rm
M_\odot}$ is $^4{\rm He}$.  We briefly describe the codes below, two of
which are grid-based, one using AMR and one not, and the last code is a
smooth particle hydrodynamics (SPH) code.

We are in particular interested in how much $^{16}{\mathrm O}$ is being
dredged up from the accretor into the SOF or further out in the star. 
In the simulations
that formed a SOF \citep[we found those with $q\lesssim 0.7$
formed a SOF in][]{staff12}, the SOF sits on top of  the core.
The SOF exists at densities of $\sim10^4-10^5~{\rm g~cm^{-3}}$.
For the purpose of estimating how much $^{16}{\mathrm O}$ has
been dredged-up to the SOF or outside of it, experience shows the core
boundary to be located at densities between
$10^5~{\rm g~cm^{-3}}$ and $10^{5.2}~{\rm g~cm^{-3}}$. Furthermore,
we also require the core to be at a temperature $T<10^8~{\rm K}$. Everything
else is therefore in the SOF or further outside, and we estimate how much
$^{16}{\mathrm O}$ exists there. This method likely underestimates the size
of  the core, and therefore overestimates the amount of $^{16}{\mathrm O}$
that has been dredged up from the core. To get a better handle on this,
we use two different density limits to define the core. 
More elaborate and refined methods may be
used to identify the core, however this will not significantly change the
amount of $^{16}{\mathrm O}$ that we find has been dredged up from the core.
In the grid-based simulations, we calculate the mass of $^{16}{\mathrm O}$
below the density limits by multiplying the mass of the gas in a cell with
the mass fraction associated with the CO and dividing by 2 (since we assume 
half the CO mass to be $^{16}{\mathrm O}$), then summing over all cells with
density below the density limits. In the SPH simulations we likewise sum the
$^{16}{\mathrm O}$ mass in all SPH particles below the density limits to
find the mass of $^{16}{\mathrm O}$ below the density limits.
In the SPH code, we keep track of the amount of $^{16}{\mathrm O}$ in each
SPH particle. In the grid-based codes, we advect two different mass
fractions. One of these is assigned to the CO, the other to the He. We will
assume that 50\% (by mass) of the CO mass fraction is $^{16}{\mathrm O}$.

The equation of state is that of a zero-temperature Fermi gas of electrons, and an ideal gas of ions.
Initially, the WDs in all the simulations are assumed to have zero temperature.
Heating can occur through shocks or adiabatic
compression.

The temperature is calculated as in \citet{staff12}:
\begin{equation}
T=\frac{E_{\rm gas}}{\rho c_v},
\label{eq:T}
\end{equation}
where $E_{\rm gas}$ is the gas internal energy and $c_v$ is the specific 
heat capacity at constant volume
\citep{segretain97}
given by:
\begin{equation}
c_v=\frac{(<Z>+1)k_B}{<A>m_H(\gamma-1)}=1.24\times10^8{\rm
ergs~g^{-1}~K^{-1}}=
~\frac{(<Z>+1)}{<A>}=6.2\times10^7 {\rm ergs~g^{-1}~K^{-1}}
\label{eq:cv}
\end{equation}
$k_B$ is Bolzmann's
constant and $m_H$ is the mass of the hydrogen atom, and we have assumed 
that $(<Z>+1)/<A>=0.5$, with $<Z>$ and $<A>$ being the
average charge and mass for a fully ionized gas. This is approximately
correct for a CO mixture, but it is an overestimate when He is present.

\subsection{Fixed grid simulations}

The fixed grid hydrodynamics code used in this work is the same as that used in
\citet{staff12}, and an earlier version of the code was described in
\citet{motl07} and \citet{dsouza06}.  In fact, the $q=0.7$ simulation is the
same as in \citet{staff12}, and we include it to compare the codes with our
previous work. This hydrodynamics code uses a
cylindrical grid, with equal spacing between the grid cells in the radial
and vertical directions.  
The resolution is $(r,z,\phi)=(226,146,256)$ cells for most of the simulations.
In addition, we ran the $q=0.5$ hybrid simulation at a higher resolution of
$(r,z,\phi)=(354,226,512)$ cells to test if the resolution plays a role in
the results. We found good agreement between the two resolutions (see section~\ref{q05subsection} for details), and
in order to conserve computer resources we therefore used the lower
resolution for the other simulations. The physical size of the grid is: $r: 0-6.1\times10^9~{\rm
cm}$, $z: -2.0\times10^9~{\rm cm}\text{ to }2.0\times10^9~{\rm cm}$. The simulations
are full 3D, so the $\phi$ direction covers $2\pi$.

The initial setup for the non-AMR simulations were also made in the same way
as in \citet{staff12}, using a self consistent field code \citep{even09}. 
Two synchronously rotating WDs are constructed, so that the donor almost
fills its Roche Lobe.  We then artificially remove orbital angular momentum
from the system at a rate of $1\%$ per orbit for a couple of orbits to force
the stars into contact faster.  The exact duration of the artificial angular
momentum removal does not appear to affect the results \citep{motl16}.

In the grid based simulations, we found that much CO material was found outside of the core, even at early times before the merger.
Speculating that this might
be a numerical artifact, we decided to ``reset'' the mass fractions in the 
hybrid accretor in the
$q=0.5$ non-AMR grid-based simulation shortly before the merger, to enforce
that all the CO material is in the core.
We thereby ignored all CO material that had so
far been dredged up, ie.  we assumed all material outside of the core of the accretor was helium shortly before the merger.

\subsection{AMR code}

The LSU code, Octo-tiger, a 3-D, finite-volume adaptive mesh refinement
(AMR) hydrodynamics code with Newtonian gravity, is a successor to previous
LSU hydrodynamics codes 
\citep{lindblom01,ott05,dsouza06,motl07,kadam16,motl16}. 
Octo-tiger
decomposes the spatial domain into a variable depth octree structure, with
each octree node containing a single Cartesian 12 X 12 X 12 subgrid. The
hydrodynamic variables are evolved using the central scheme of 
\citet{kurganov00}, while the gravitational field is computed using the fast
multipole method (FMM) presented by \citet{dehnen00}. The AMR simulations
presented here use 8 levels of refinement. These simulations were also
presented in \citet{montiel15} in order to study mass loss from WD mergers.

\subsubsection{Initial Conditions}

To generate our initial models for the AMR simulations, we used a method 
similar to the self
consistent field (SCF) technique described by \citet{even09}. The SCF method
solves the hydrostatic balance equation in the presence of gravity,
\begin{equation}
h + \Psi = \Psi_0,
\end{equation}
where $\Psi_0$ is a constant unique to each star. The isentropic enthalpy,
$h$, is defined as 
\begin{equation}
\label{hbal}
h\left[\rho\right] = \int_0^{P=P\left[\rho\right]} \frac{dP'}{\rho'}.
\end{equation}
For the zero temperature WD equation of state, 
\begin{equation} 
h\left[\rho\right] = \frac{8 A}{B}\sqrt{
\left(\frac{\rho}{B}\right)^\frac{2}{3} + 1} .
\end{equation}

\citet{even09} require the choosing of two boundary points for the donor
star, each on the line of centers between the stars and on opposite sides of
the donor. For our initial model,
instead of fixing the boundary point closest to the accretor in space, we
define it to be the L1 Lagrange point. This sets $\Psi_0$ for the accretor
to $\Psi_{L1} + h\left[0\right]$, where $\Psi_{L1}$ is the effective
potential at the L1 point. This ensures the donor Roche lobe is filled. The
donor WD is taken to be 100 \% helium. For hybrid accretor models, the core is taken to contain an evenly distributed mixture of equal parts of carbon and oxygen, while the envelope is 100 \% helium. The non-hybrid accretor contains equal parts carbon and oxygen throughout.

Because the discretization used for the initial conditions and the
discretization that results from writing the time invariant version of the
semi-discrete evolution equations are not exactly the same, the initial
model is not in exact equilibrium when it begins evolving. As a result, the
outer edges of each star diffuse slightly at the very beginning of the
simulation. In the case of the donor, this causes Roche lobe overflow,
leading to mass transfer.

\subsection{SPH code}

Smoothed particle hydrodynamic (SPH) merger calculations were carried out in
SNSPH \citep{fryer06}.  The code contains a highly scalable hashed
oct-tree data structure \citep{warren95} to support efficient neighbor
finding and is coupled with a multipole expansion to calculate the
gravitational potential and
was run with a traditional SPH scheme with the cold WD EOS and
standard ($\alpha = 2, \beta = 1$) artificial viscosity parameters.

Our SPH simulations contain the same initial data as the grid-based
calculations, being generated from the self-consistent-field (SCF) code.
Initial particle distributions were prepared with the
method in \citet{diehl15} which converts cylindrical grid-based data into a
particle representation and iteratively re-arranges particles to recover a
distribution reminiscent of Weighted Voronoi Tessellation (WVT) initial
conditions.  Setups for each pair of stars contained $\sim 20$M particles of
nearly equal mass\footnote{for a simulation with 20 million particles and 
$M_{\rm tot}=0.7~{\rm M_\odot}$,
each particle has a mass of
$\approx7\times10^{25}~{\rm g}=3.5\times10^{-8}~{\rm M_\odot}$.}. 
Each star in the WD binary system was subsequently allowed to relax
individually for a short period of time prior to the merger calculation to
prevent stellar oscillations and premature WD heating.

Merger calculations were run on 256 cores and required approximately 3 weeks
($\sim100,000$ CPU hours) to run to completion.  White dwarf binaries were
allowed to orbit about a half-dozen times before the stars were driven into
contact by removing a small amount ($1\%$) of the system's angular momentum
during each orbit.  Once mass transfer was initiated, angular momentum was
no longer extracted from the system.  Calculations were run out to $\sim 5$
orbital periods post-merger.

\subsection{Formation of hybrid WDs}

The scenario forming a hybrid WD in a short period binary with a He WD
is complex and therefore it is valuable to review it in detail here to show that such objects can form, and that they do so in binaries. \citet{rappaport09} outlined how this may happen, and
here we just briefly summarize the scenario they discussed for Regulus: A
binary system consists of two main sequence stars in a binary with an orbital
period of $\sim40$ hours and with masses $2.1~{\rm
M_\odot}$ for the primary and $1.74~{\rm M_\odot}$ for the companion, that
eventually will become the hybrid WD. The more massive primary
evolves off the main sequence first, transferring its envelope mass to the
companion, which also causes the period to expand to $\sim40~{\rm days}$,
which corresponds to a separation of $\sim76~{\rm R_\odot}$. This way, 
a $\sim3.4~{\rm M_\odot}$ star is orbited by a $\sim0.3~{\rm M_\odot}$ He
WD. It is crucial that the separation is of this size, since that will cause
the $3.4~{\rm M_\odot}$ star to overflow its Roche lobe when it is near the
tip of the red giant branch. At this point it has a He core of roughly
$0.48~{\rm M_\odot}$. The mass transfer leads to a common envelope, and if
this interaction ejects the envelope before the helium core merges with the
He WD, a short period binary results consisting of the He WD and the helium
core. The helium core is sufficiently massive that it will ignite, and
\citet{rappaport09} found that the period should be larger than about 80
minutes or else the helium star would overflow its Roche lobe while still
burning helium in the center. Once core helium burning ceases, the star
will cool and contract to form a hybrid He/CO WD. Emission of gravitational
waves will bring the WDs into contact, with the bigger, less massive He WD
as the donor. Other masses than those
discussed in \citet{rappaport09} can likely also lead to a binary consisting
of a hybrid He/CO WD and a He WD, with the required periods adjusted
accordingly. However, since some fine-tuning is
needed in order to get the correct separation following both the initial 
mass transfer that created the He WD, and following the common envelope
interaction that exposed the He core, it is likely that this is a rare
process. 

The formation rate of RCB stars is not known. Assuming that RCB stars are formed by the merger of a CO and an He WD, \citet{karakas15} found an RCB birthrate of $1.8\times10^{-3}~{\rm yr}^{-1}$. 
\citet{brown16} found a similar merger rate for CO+He WDs.
Depending on the RCB lifetime this can mean that there are a few hundred RCB stars in the galaxy, in agreement with estimates in \citet{lawson90}, but less than the several thousands estimated in \citet{han98}. In any case, considerable uncertainty surrounds the number of RCB stars in the galaxy, their lifetime, and hence their birthrate. It is therefore not unthinkable that a rare process like the merger of a hybrid He/CO WD with a CO WD could be the formation channel of RCB stars.

\section{Results}
\label{resultssection}

\subsection{Non-hybrid accretor, $M_{\rm tot}\approx 0.9 M_\odot$, $q=0.7$}

\begin{figure} 
\includegraphics[width=0.85\textwidth]{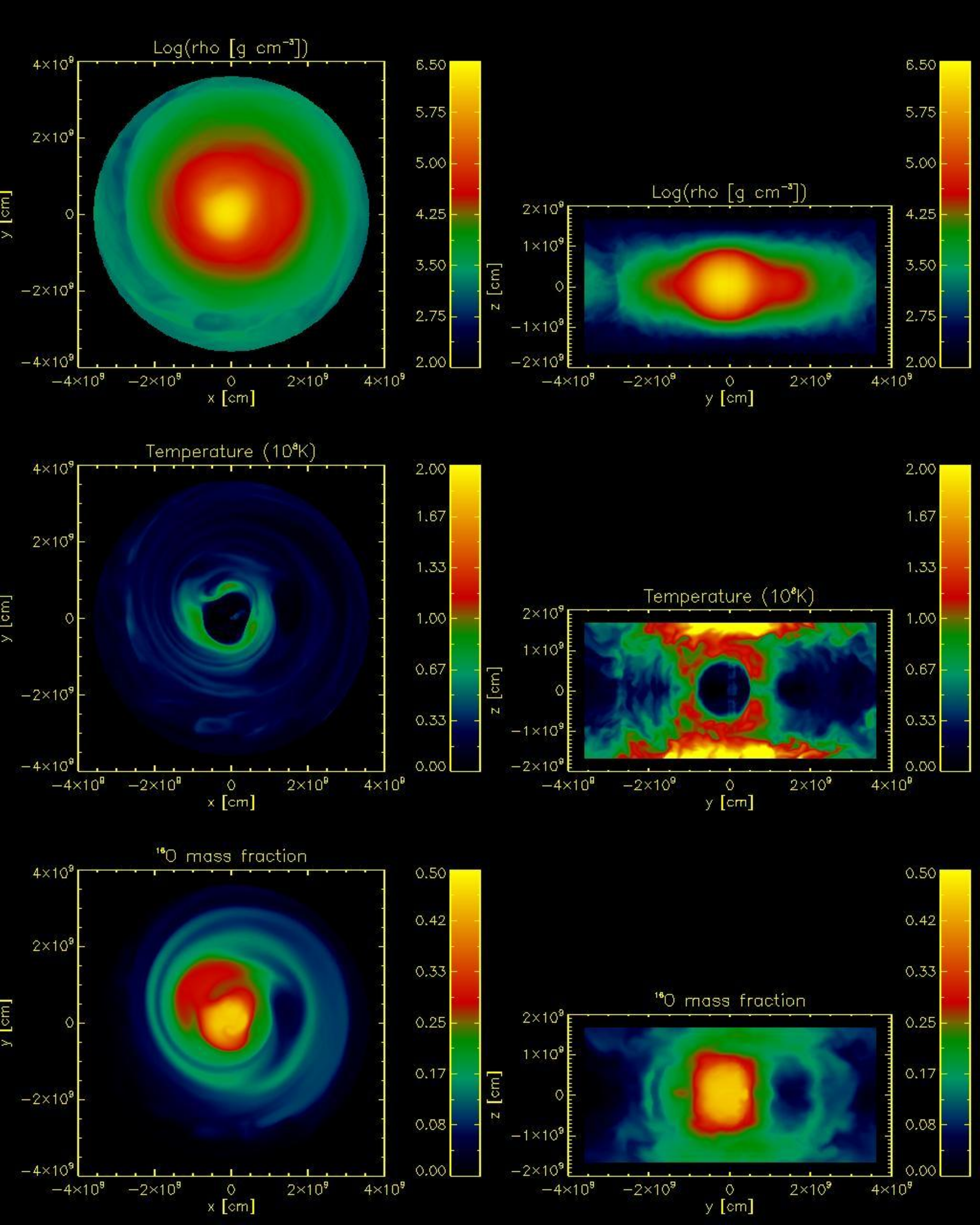}
\caption{The results of the $q=0.7$ non-AMR grid simulation \citep[this
simulation was also presented in][]{staff12} showing equatorial slices
in the left column and slices perpendicular to this in the right column. The
top row shows logarithm
of density, the middle row shows temperature, and the bottom row shows the
accretor mass fraction. The blob is clearly visible to the right of the core
in the equatorial plots. The perpendicular slices in the right column 
stretch from the center of
the grid to $3.6\times10^9~{\rm cm}$, from $-1.7\times10^9$ to
$1.7\times10^9~{\rm cm}$ in the vertical direction, and are made through the blob.}
\label{jan07} 
\end{figure}

\begin{figure} 
\includegraphics[width=\textwidth]{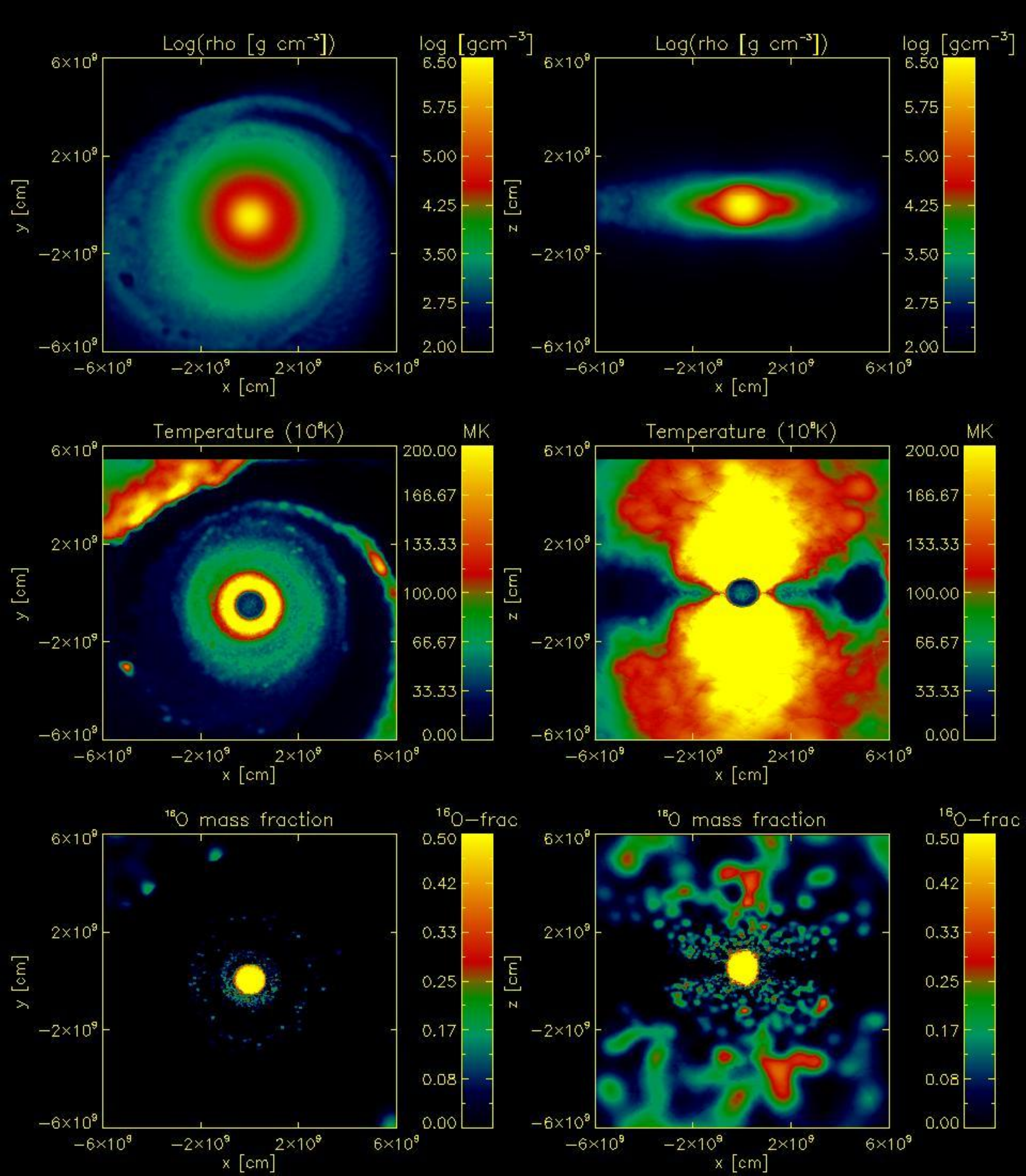}
\caption{The results of the $q=0.7$ non-hybrid SPH simulation, showing 
equatorial slices
in the left column and slices perpendicular to this through the middle of
the core in the right column. The
top row shows logarithm
of density, the middle row shows temperature, and the bottom row shows the
$^{16}{\mathrm O}$ mass fraction. A weak blob-like feature is visible to the right of the merged core in the equatorial plots, and the vertical slices are made through this.}
\label{brandon07}
\end{figure}

\begin{figure} 
\includegraphics[width=0.9\textwidth]{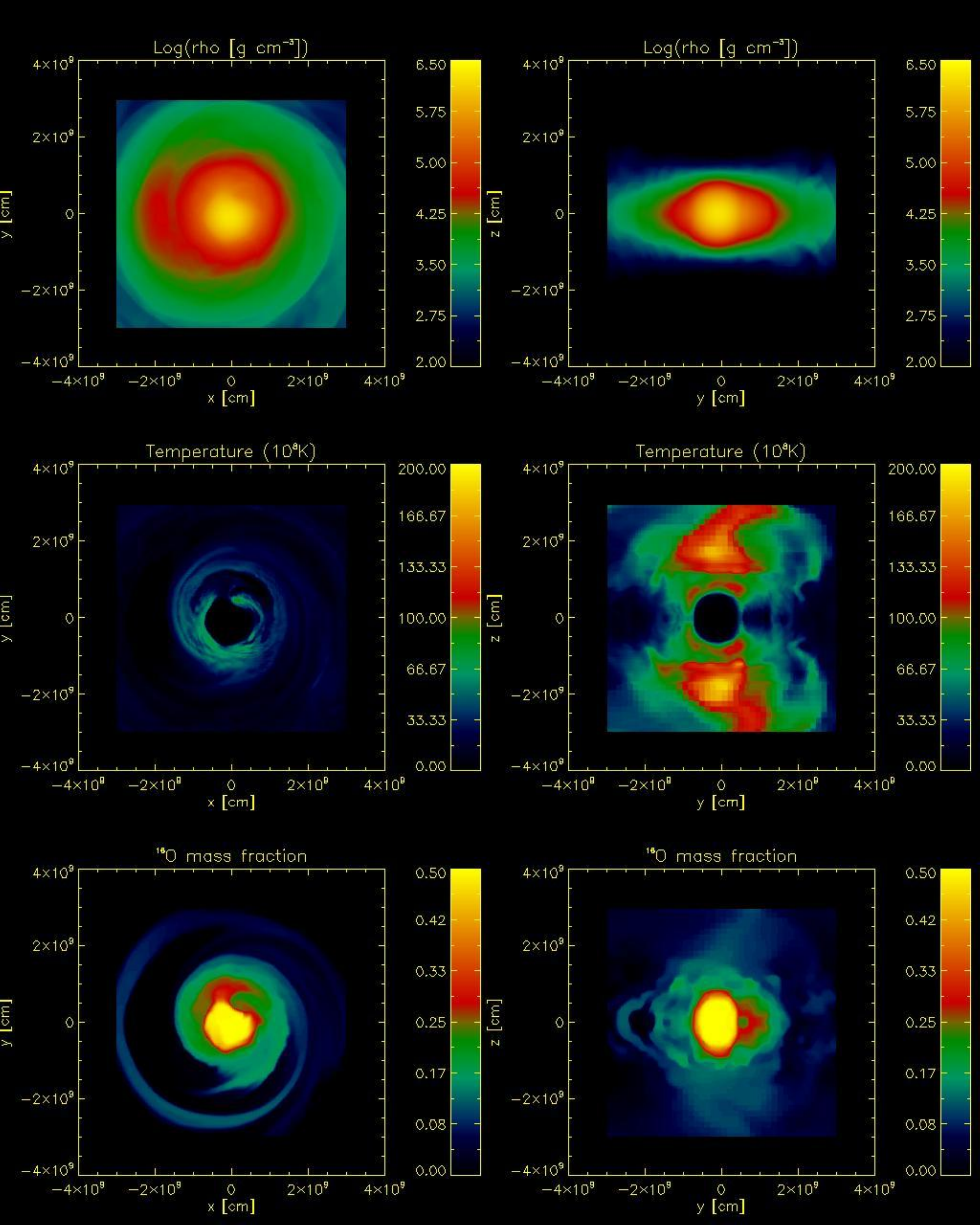}
\caption{The results of the $q=0.7$ AMR grid simulation. {\it Upper panel: }
Logarithm of density, {\it middle panel: } temperature, {\it lower panel:
}CO mass fraction. The left image in each panel shows the equatorial slice,
while the right image shows a slice perpendicular to it, through the middle
of the grid, taken through the blob, which is visible to the left of the merged core.}
\label{dominic07}
\end{figure}

The results from the grid-based, non-AMR simulations of the $q=0.7$, $M_{\rm
tot}=0.9~{\rm M_\odot}$ setup from \citet{staff12} are shown again here in
an equatorial slice and a 
slice perpendicular to the orbital plane in Fig.~\ref{jan07} \citep[the time of
the snapshot is slightly different from that in][]{staff12}.  We find
that there is $0.09 M_\odot$ of accretor material at 
densities $\rho<10^{5.2}~{\rm g~cm^{-3}}$ following the merger, or
$0.07~{\rm M_\odot}$ of accretor material at densities
$\rho<10^{5}~{\rm g~cm^{-3}}$.  
If we assume half of this to be $^{16}{\mathrm O}$, we get that there is
$0.035-0.045~{\rm M_\odot}$ of $^{16}{\mathrm O}$ outside of the merged core
immediately following the merger. The amount of $^{16}{\mathrm O}$ below
these densities grows following the merger and we find $0.055-0.07~{\rm M_\odot}$
of $^{16}{\mathrm O}$ below these densities at the end of the
simulation. We find temperatures up to $1.5\times10^8~{\rm K}$ in the SOF.
The maximum density in the merged core is $\sim10^6~{\rm g~cm^{-3}}$.

Figure~\ref{brandon07} shows
the density, temperature, and $^{16}{\mathrm O}$  mass fraction of the $q=0.7$ non-hybrid SPH simulation. 
Very little of the $^{16}\mathrm{O}$ that is dredged up from the accretor
ends up in the equatorial plane.
Most of the helium from the donor is found in the equatorial plane.  
We find $\sim0.01~{\rm M_\odot}$ of $^{16}{\mathrm O}$ at
densities $\rho<10^{5}~{\rm g~cm^{-3}}$. Likewise, we find
$\sim0.02~{\rm M_\odot}$ of $^{16}{\mathrm O}$ at
densities $\rho<10^{5.2}~{\rm g~cm^{-3}}$. 
We find temperatures above $2.5\times10^8~{\rm K}$ in the SOF in this SPH 
simulation, and densities $3\times10^6~{\rm g~cm^{-3}}$ in the merged core. 

In the $q=0.7$ AMR simulations (Fig.~\ref{dominic07}) we find about $0.055~{\rm M_\odot}$
($1.1\times10^{32}~{\rm g}$) of CO material at densities $\rho<10^5~{\rm
g~cm^{-3}}$, and about $0.075~{\rm M_\odot}$ ($1.5\times10^{32}~{\rm g}$) of
CO material at densities $\rho<10^{5.2} ~{\rm g~cm^{-3}}$ following the
merger.  Again assuming half of this is $^{16}{\mathrm O}$, we get that
roughly $0.03-0.04~{\rm M_\odot}$ of $^{16}{\mathrm O}$ is outside of the
core following the merger.  This is quite similar to what we found in the
non-AMR simulations.  It is about a factor 2 more than what we found in the
SPH simulations. We find that the temperatures barely reach
$1\times10^8~{\rm K}$ in the SOF, and densities in the merged core 
reach $2\times10^6~{\rm g~cm^{-3}}$.

\begin{figure}  
\includegraphics[width=0.5\textwidth]{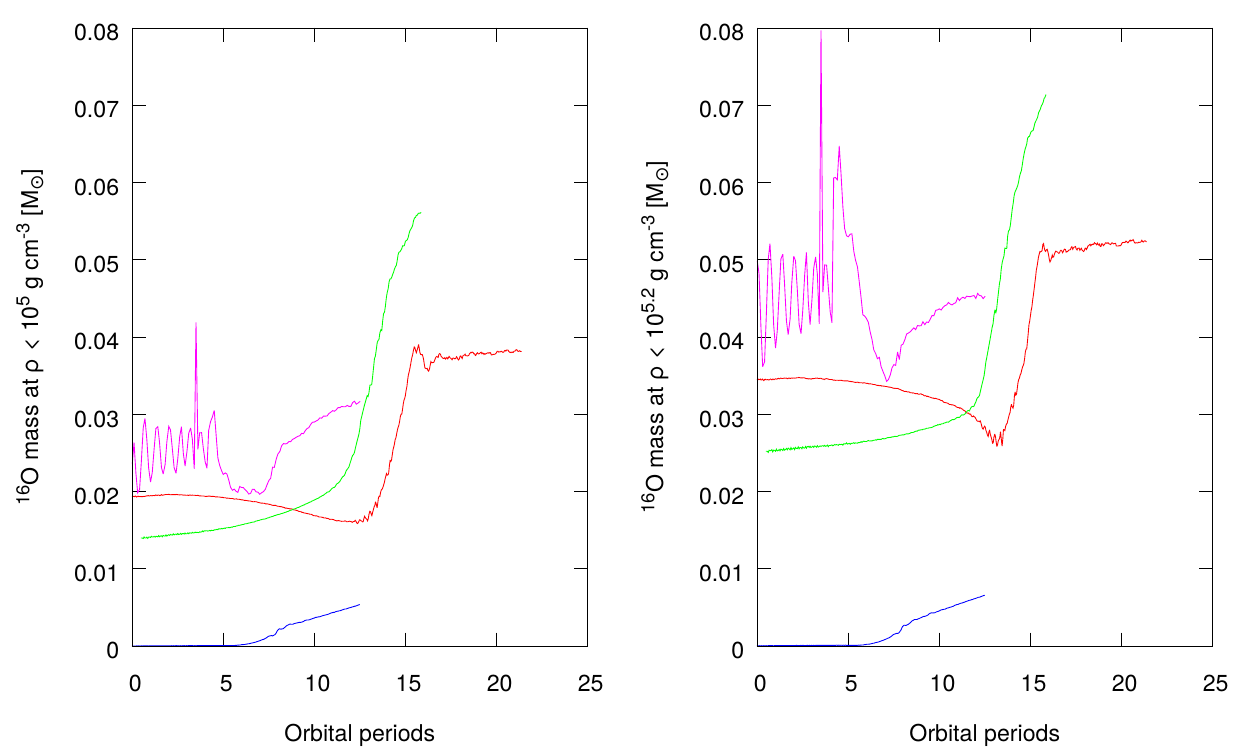}
\caption{The mass of oxygen at densities below $\rho<10^5~{\rm
g~cm^{-3}}$ (left panel) and $\rho<10^{5.2}~{\rm g~cm^{-3}}$ (right panel) as
a function of time in the $q=0.7$ non-AMR simulation (green curve), 
AMR simulation (red curve), SPH simulation (blue curve), and SPH simulation mapped to a grid (purple curve).}
\label{q07amrcofrac}
\end{figure}

In Fig.~\ref{q07amrcofrac} we show the mass of CO material at densities
below $\rho<10^5~{\rm g~cm^{-3}}$ and $\rho<10^{5.2}~{\rm g~cm^{-3}}$ as a
function of time in all of the simulations.  As this is not a hybrid simulation,
even initially there is CO material at these relatively low densities in the grid-based simulations, but not in the SPH simulation, showing that the density structure is slightly different between the simulations. We note that when we transform the SPH data to a grid, we also find similar amount of CO material at these lower densities.
Interestingly, during the merger the CO core gets squeezed sufficiently that
the amount of CO material below these densities drops in the AMR simulation. 
We do not see this squeezing in the non-AMR simulations or the SPH
simulations.
However, the merger leads to dredge up, and following
the merger there is $50-80\%$ more CO material at lower densities than
initially.

\subsection{Hybrid accretor, $M_{\rm tot}=0.71 M_\odot$, $q=0.5$}
\label{q05subsection}

\begin{figure}   
\includegraphics[width=0.85\textwidth]{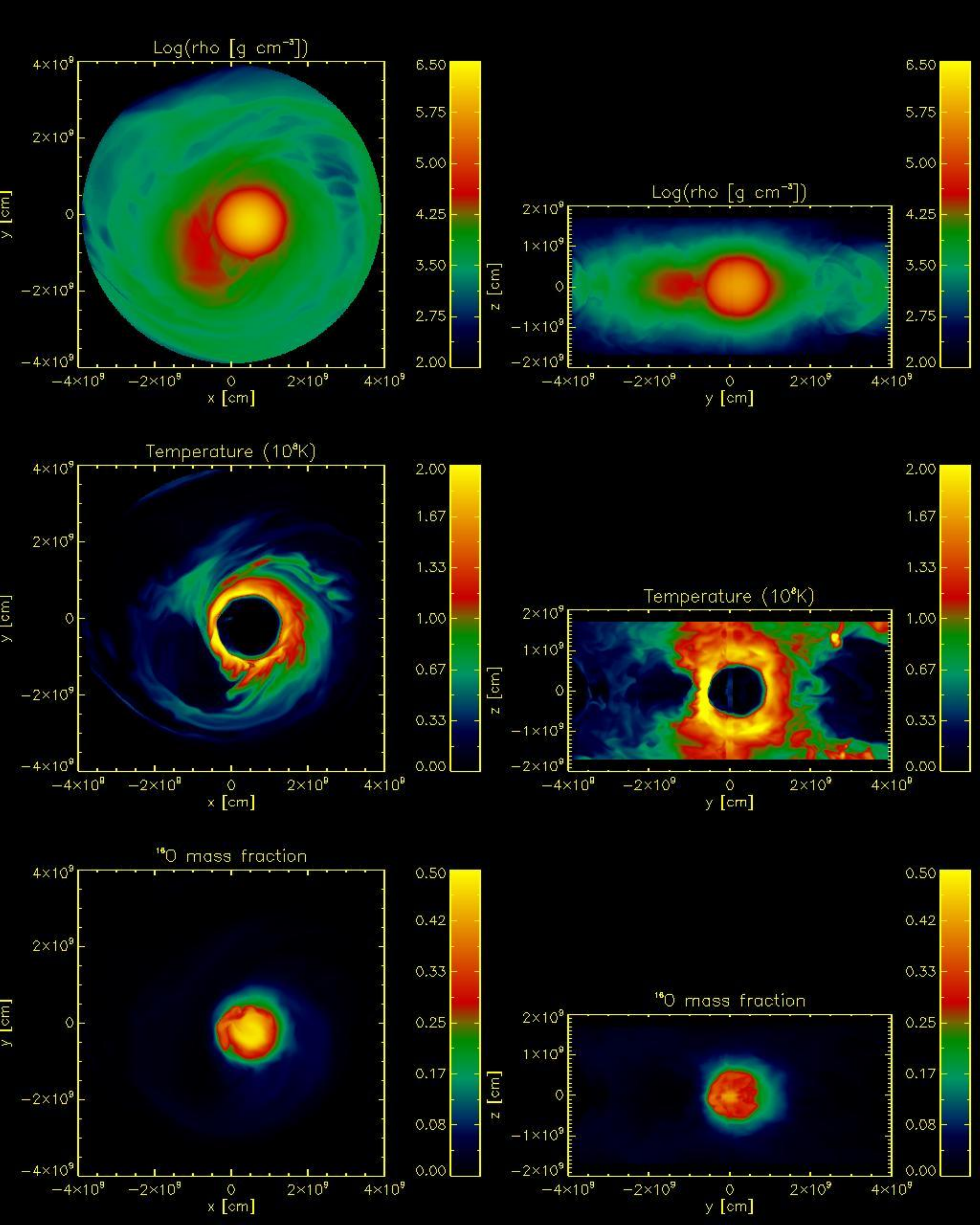}
\caption{High resolution non-AMR grid-based hybrid simulation with $q=0.5$, showing equatorial slices
in the left column and slices perpendicular to this in the right column. The
top row shows logarithm
of density, the middle row shows temperature, and the bottom row shows the
accretor mass fraction. The blob is clearly visible to the left of the core
in the equatorial plots. The perpendicular slices in the right column 
stretches from the center of
the grid to $5\times10^9~{\rm cm}$ in the radial direction, and from
$-2\times10^9$ to $2\times10^9~{\rm cm}$ in the vertical direction, and it is made through the blob.}
\label{jan05hybrid}
\end{figure}

In the high resolution non-AMR grid simulation, we find that
$\sim0.04 M_\odot$ accretor material is at
$\rho<10^{5.2}~{\rm g/cm^3}$ ($\sim 0.02 M_\odot$ at $\rho<10^5~{\rm
g/cm^3}$). 
If we assume that the accretor consists of equal amounts of
$^{16}\mathrm{O}$ and $^{12}\mathrm{C}$, then half of this is
$^{16}\mathrm{O}$. We also find that this amount keeps increasing with time
past the merger event, indicating that there is some artificial diffusion of
the mass fractions in our non-AMR grid-based simulations. In the merged
core, we find densities of $\sim1.7\times10^6~{\rm g/cm^3}$, and temperatures in
the SOF reaches $1.6\times10^8~{\rm K}$ in the high resolution simulation. 

For comparison, the amount of accretor mass below the threshold densities are $0.05~{\rm M_\odot}$ for $\rho<10^{5.2}~{\rm g/cm^3}$ and $\sim 0.03 M_\odot$ at $\rho<10^5~{\rm g/cm^3}$ for the lower resolution simulation. 
In the lower resolution simulations we also found densities up to $\sim1.5\times10^6~{\rm g/cm^3}$ and sustained temperatures in the SOF up to $\sim1.8\times10^8~{\rm K}$.
Using the amount of accretor material at densities below a certain threshold to compare resolutions is difficult, since the amount of accretor material keeps increasing with time following the merger (see Fig.~\ref{q07amrcofrac}). 
Since the merger is a drawn out process, that alone can not be used as a measure of time. 
The accretor masses quoted here are therefore found 1.5 orbits after the last frame that showed a density maximum for the donor. 
Using the temperature to compare simulations with different resolutions is also inaccurate, since the temperature can fluctuate quite a bit. 
The temperatures quoted are therefore temperatures that we found could be sustained for some time in the SOF, but the exact value is somewhat subjective.
The central density is resolution dependent, since a higher resolution better resolves also the central region, where the density increases towards the center.
Nevertheless, with all this in mind we judge that there is reasonably good agreement between the simulations with different resolution.

The accretor material outside of the core is predominantly located in or
around the equatorial plane (see Fig.~\ref{jan05hybrid}).
It encapsulates the blob (see below), and very
little accretor material is being pushed vertically in this
simulation. The majority of the dredge-up has happened in or around the
equatorial plane.

As in all the other grid-based non-AMR simulations that we have performed,
we find a donor-material rich, cold blob forming outside of the high density
core (Fig.~\ref{jan05hybrid}). This blob appears to be some of the last of
the donor material to fall onto the accretor, but it manages to stay
together and not diffuse out. Over time, however, its size does decrease, perhaps
in part  because of the artificial diffusion of the accretor and donor mass
fractions.

\begin{figure}   
\includegraphics[width=0.95\textwidth]{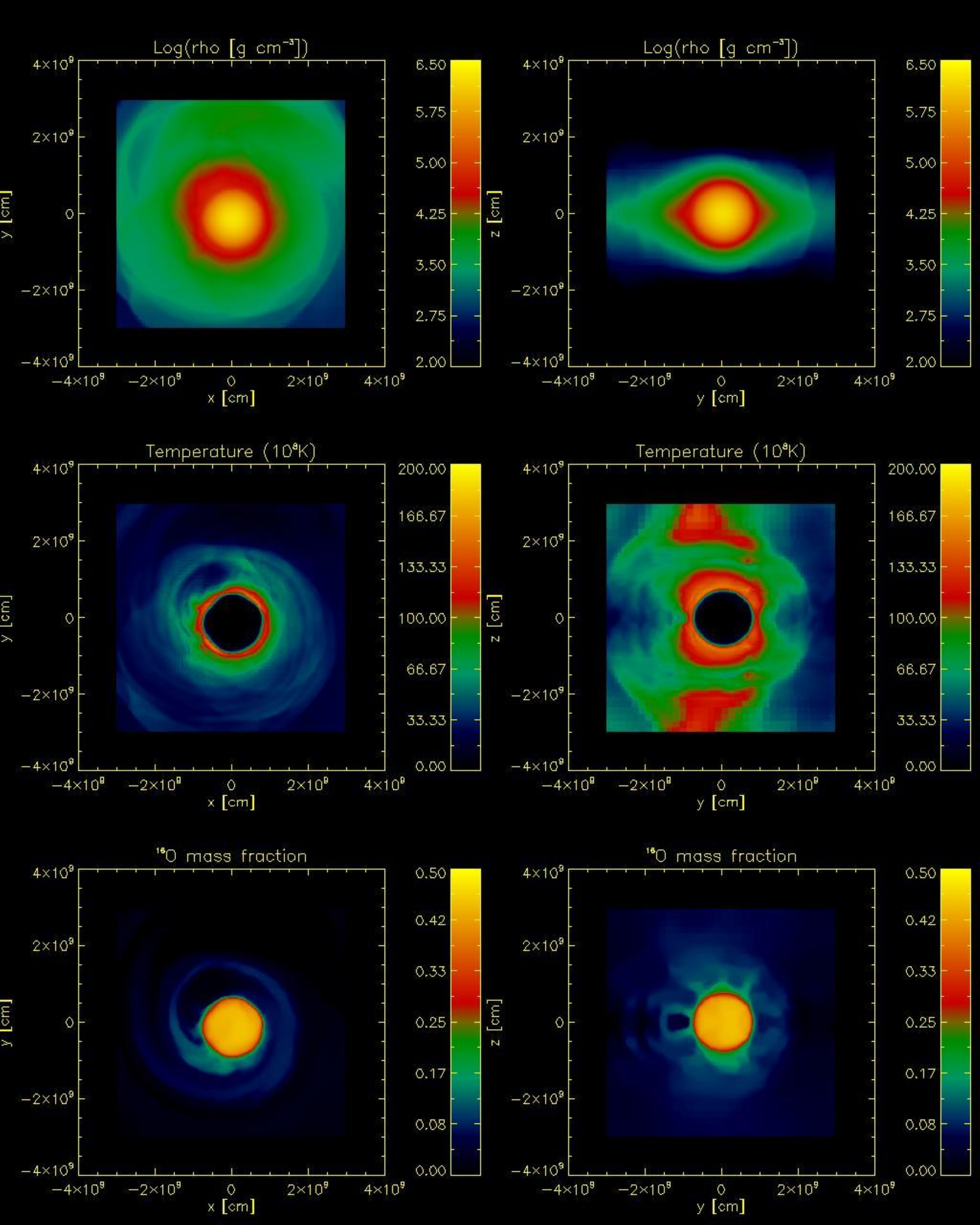}
\caption{Log density (top panel), temperature (middle panel), and CO mass
fraction (bottom panel) for
the $q=0.5$ hybrid simulation with the AMR grid code, taken in the equatorial
slice (left frames), and perpendicular to it (right frames). The blob is visible to the right and above the merged core. The perpendicular slices are made through the lower part of the blob at $y=0~{\rm cm}$.}
\label{dominic05hybridcomp}
\end{figure}

In Fig.~\ref{dominic05hybridcomp} we show the results of the AMR simulation
for the hybrid simulation with $q=0.5$ and total mass of $0.758~{\rm
M_\odot}$. Again, as in the other grid-based simulations, we find a
cold, higher density, donor rich blob outside of the merged core (sitting at
the upper left of the core in Fig.~\ref{dominic05hybridcomp}). In fact, the
AMR simulation looks very similar to the non-AMR simulation in many ways.
There are ``fingers'' of donor material ``attacking'' the core before and
during the mergers. However, contrary to the non-AMR simulation, this does
not lead to much contamination of the CO core, which maintain a high CO
fraction of about $0.9$ after the merger.

We find
densities up to $2\times10^6~{\rm g/cm^3}$.
We again find a SOF surrounding the
core, with temperatures up to $2\times10^8~{\rm K}$. 
As in the
non-AMR simulations, the SOF contains a near equal mix of accretor and donor
material.

In the AMR grid simulation, we find that the amount of accretor material at
densities  $\rho<10^5~{\rm g/cm^3}$ increases from $0.02~{\rm M_\odot}$
immediately before the merger to $0.06~{\rm M_\odot}$ at the end of the
simulation, while at densities below $\rho<10^{5.2}~{\rm g/cm^3}$ it climbs
from about $0.04~{\rm M_\odot}$ immediately before the merger to about
$0.07~{\rm M_\odot}$ at the end of the simulation.
If we again assume that the
accretor material consists of equal amounts of $^{16}\mathrm{O}$ and
$^{12}\mathrm{C}$, then half of this is
$^{16}\mathrm{O}$. We note that this number is quite close to the number
found in the non-AMR simulation (remembering that in the non-AMR simulation
the CO mass outside of these densities had been artificially set to zero
shortly before the merger), which gives us some confidence in this
result. Of interest is that as soon as mass transfer starts prior to the
merger, the CO mass at low densities rapidly grows and reaches a
plateau-value of about $0.05~{\rm M_\odot}$ for $\rho<10^5~{\rm g/cm^3}$ and
$0.09~{\rm M_\odot}$ for $\rho<10^{5.2}~{\rm g/cm^3}$. This is already much
too high to explain the oxygen ratios in RCBs.

\begin{figure}   
\includegraphics[width=\textwidth]{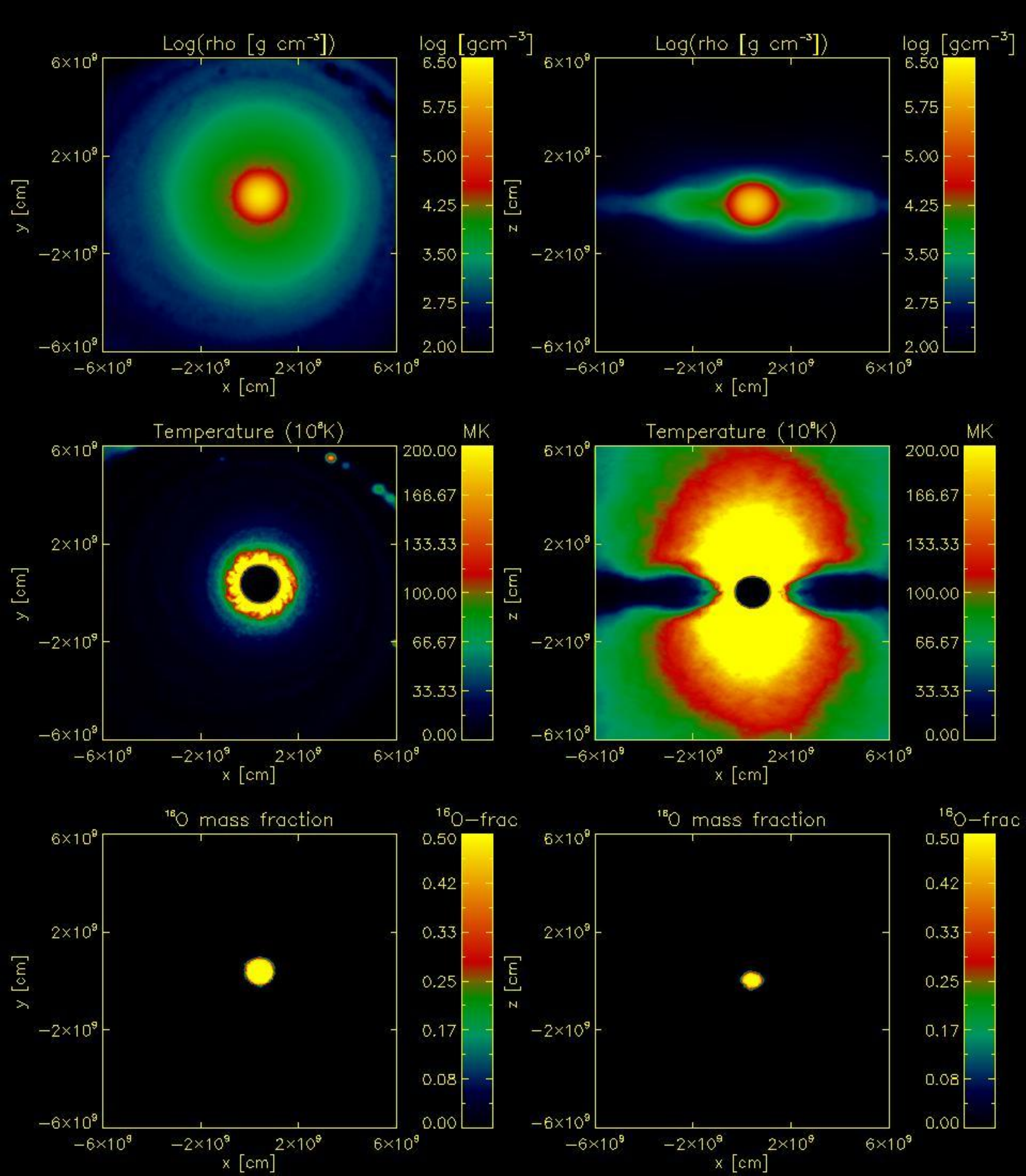}
\caption{SPH simulations with hybrid accretor and $q=0.5$ mass ratio, 
showing equatorial slices
in the left column and slices perpendicular to this through the middle of
the core in the right column. The
top row shows logarithm
of density, the middle row shows temperature, and the bottom row shows the
$^{16}{\mathrm O}$  mass fraction. The merged object is very axisymmetric, and no blob is visible. The vertical slices are made at $y=0~{\rm cm}$.}
\label{brandonq05hybrid}
\end{figure}

Figure~\ref{brandonq05hybrid} shows the logarithm of
the density, the temperature, and the $^{16}{\mathrm O}$ mass fraction, in the
equatorial plane and perpendicular to it for the SPH simulation with $q=0.5$
and a hybrid WD accretor. We find no $^{16}{\mathrm O}$
at densities below $\rho<10^{5.2}~{\rm g/cm^3}$. We find densities up to
$3\times10^6~{\rm g/cm^3}$ 
in the merged core. An SOF did form, with maximum temperature of
$\sim2.0\times10^8~{\rm K}$. 
There is no blob formed in this simulation, and indeed we see that the
resulting object is quite symmetric.

\subsection{Hybrid accretor, $M_{\rm tot}\approx 0.9 M_\odot$, $q=0.8$}

\begin{figure}   
\includegraphics[width=0.85\textwidth]{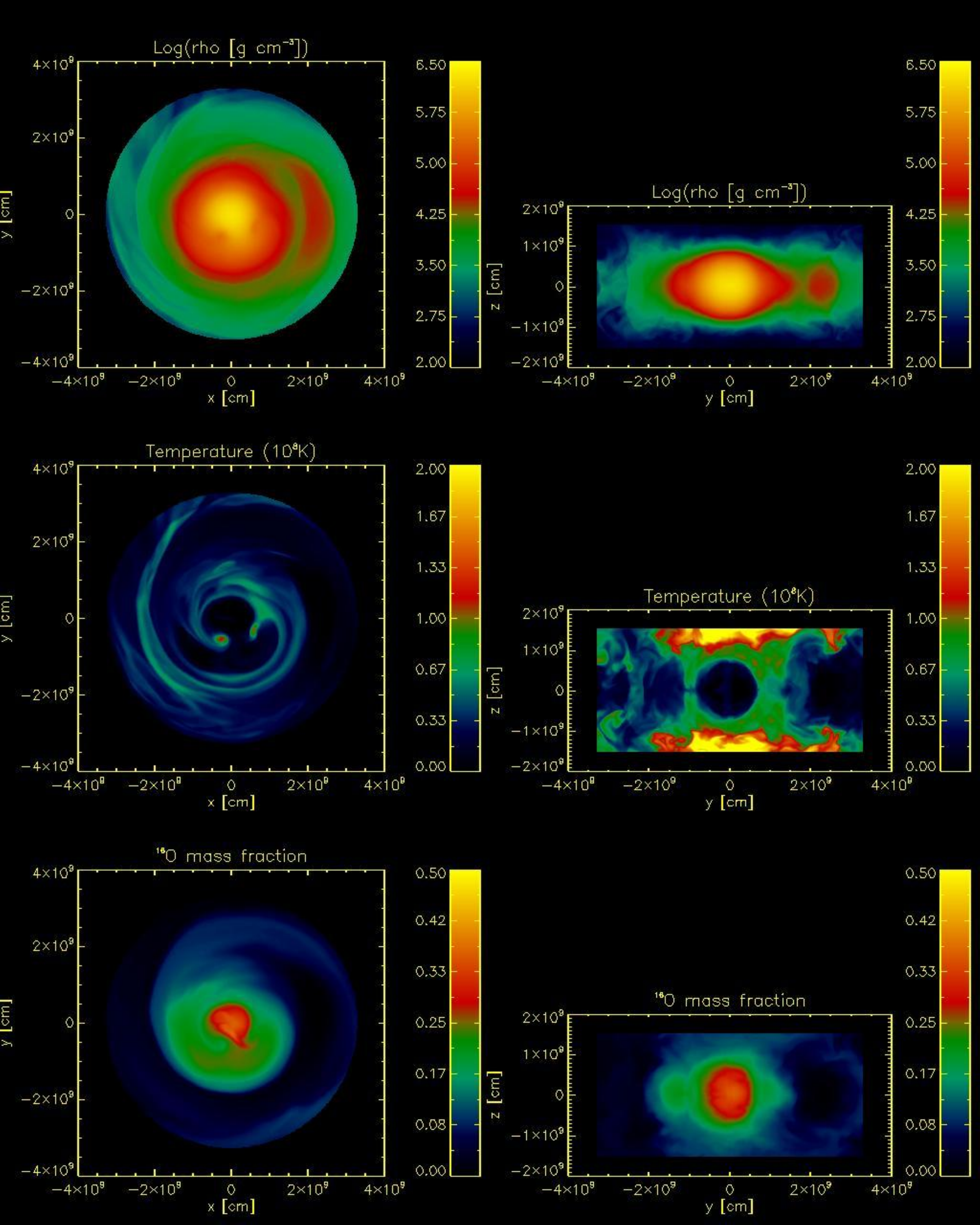}
\caption{Non-AMR grid-based high-resolution hybrid simulation with $q=0.8$, showing equatorial slices
in the left column and slices perpendicular to this in the right column. The
top row shows logarithm
of density, the middle row shows temperature, and the bottom row shows the
accretor mass fraction. The blob is visible to the right of the core
in the equatorial plots. The perpendicular slices in the right column 
stretch from the center of
the grid to $3.8\times10^9~{\rm cm}$, and are made through the blob.}
\label{jan08hybrid}
\end{figure}

In the non-AMR grid-based simulations, we find densities up to
$3.2\times10^6~{\rm g~cm^{-3}}$. We show the logarithm of the density,
the temperature, and the accretor fraction in the equatorial slice and in a
slice perpendicular to it in Fig.~\ref{jan08hybrid}.
In \citet{staff12} we found that high-q simulations do not have a SOF as
this is destroyed in the merger of the two cores, and a similar thing 
occurs in the grid-based non-AMR $q=0.8$ simulation we present here.  
The hottest regions at high densities in this simulation
are a few spots deep inside the high-density merged core that reach about
$1.25\times10^8~{\rm K}$. Surrounding the core there is, however, a hot shell
with temperatures up to $10^8~{\rm K}$, although in the equatorial plane it 
is noticeably cooler.
As a result of the violent core merger, much
$^{16}{\mathrm O}$ is dredged-up.
We find that  $0.06 M_\odot$ of accretor
material is at $\rho<10^5~{\rm g/cm^3}$ ($0.087 M_\odot$ of accretor
material at $\rho<10^{5.2}~{\rm g/cm^3}$). Again assuming that half of this is
$^{16}{\mathrm O}$, then about $0.03M_\odot$ of $^{16}{\mathrm O}$ is
outside of $\rho<10^5~{\rm g/cm^3}$. 

\begin{figure}   
\includegraphics[width=0.95\textwidth]{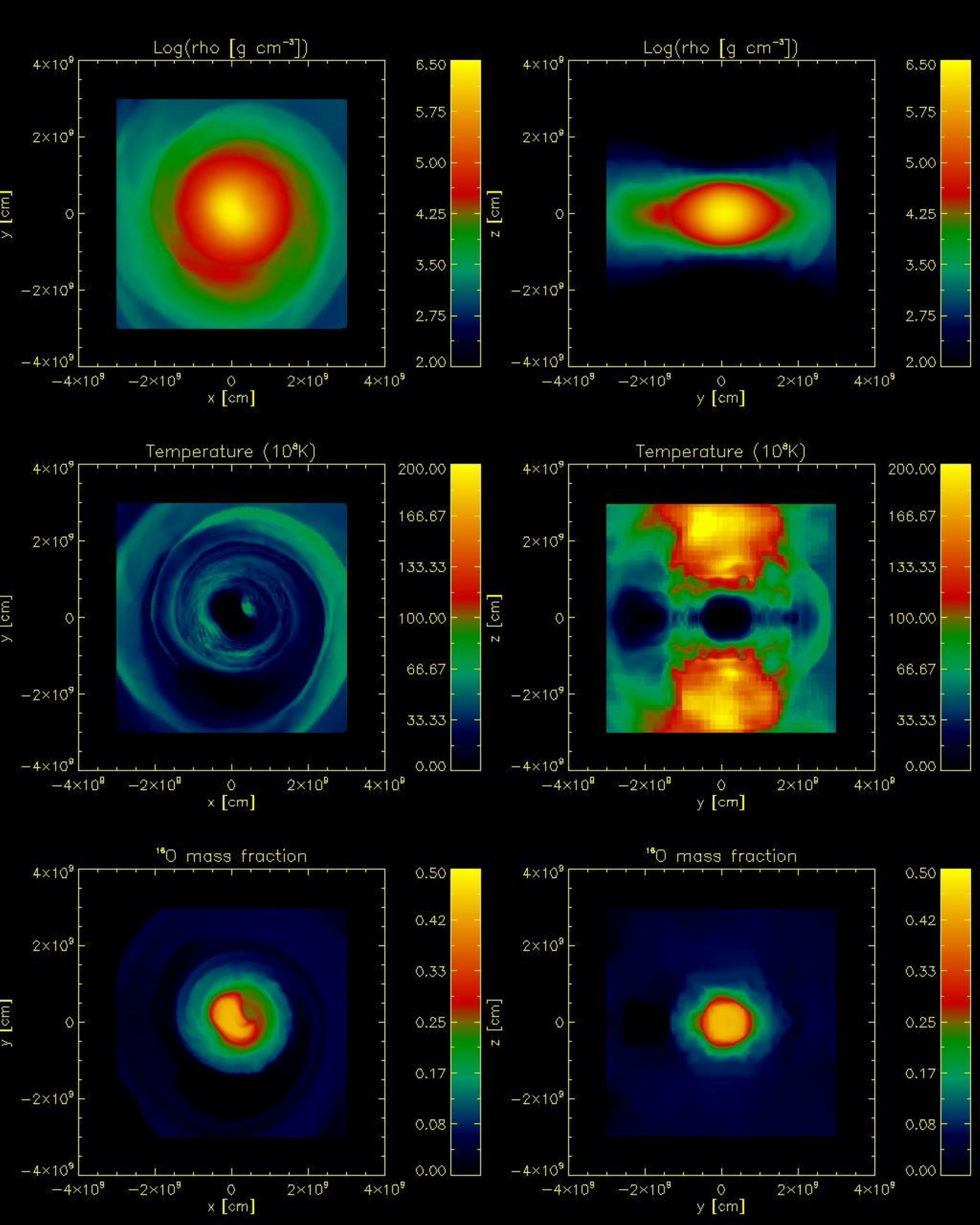}
\caption{Log density (top panel), temperature (middle panel), and CO mass
fraction (bottom panel) for
the q=0.8 hybrid simulation with the AMR grid code, taken in the equatorial
slice (left frames), and perpendicular to it (right frames). The vertical slices are made through the blob, which is visible below the merged core in the equatorial slices.}
\label{dominic08}
\end{figure}

In Fig.~\ref{dominic08} we show the logarithm of density, accretor mass
fraction, and temperature, in the equatorial slice and a slice perpendicular
to it for the grid-based AMR simulations for $q=0.8$ and with a total mass
of $1.01~{\rm M_\odot}$. The results compare well with the non-AMR grid-based
simulations. We find densities up to $5\times10^6~{\rm g~cm^{-3}}$, and
temperatures up to $8\times10^7~{\rm K}$, which is found in a ``hot spot'' in the merged core (as in the non-AMR simulation). The merged core
contains much donor material as a result of a core merger, and the accretor rich part of the core is
highly distorted and not spherical. There is no clear SOF, although there is
a region surrounding the merged core with a slightly higher temperature 
($T\sim5\times10^7~{\rm K}$), at a density of $10^4 - 10^5~{\rm g~cm^{-3}}$ in the equatorial plane. As in the 
non-AMR simulation, the equatorial plane is noticeably cooler. Also as in the
non-AMR grid-based simulation, we find a cold, donor rich blob outside of
the merged core. 

Even before the merger, much accretor mass is at densities below
the density thresholds. Following the merger, we find $0.08~{\rm M_\odot}$ of CO mass
at densities below $10^{-5}~{\rm g~cm^{-3}}$, and $0.12~{\rm M_\odot}$ at
densities below $10^{-5.2}~{\rm g~cm^{-3}}$, corresponding to
$0.04-0.06~{\rm M_\odot}$ of $^{16}{\mathrm O}$ in the SOF or outside. This
is slightly more than what we found in the non-AMR grid-based simulations.

\begin{figure}   
\includegraphics[width=\textwidth]{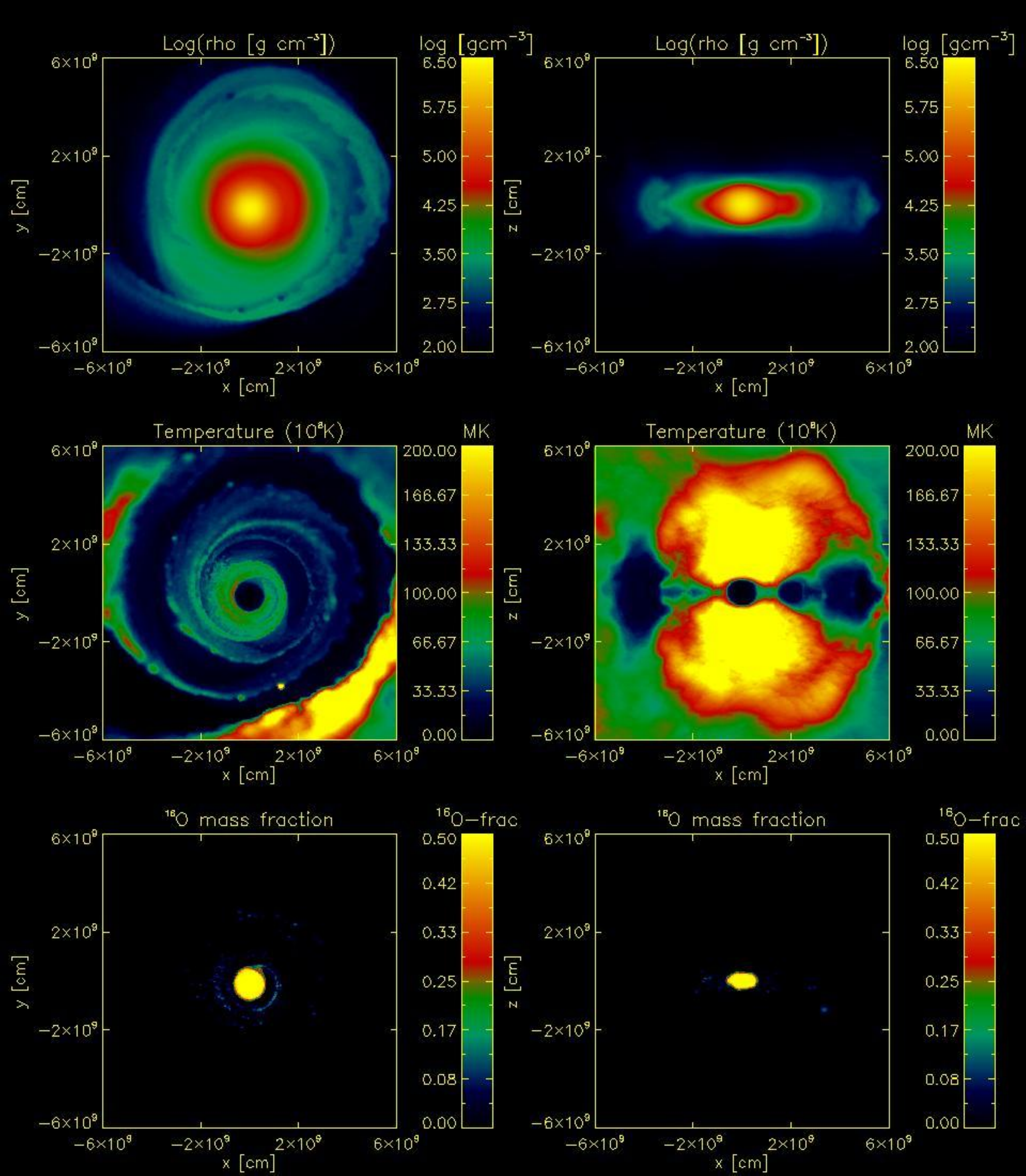}
\caption{SPH based hybrid simulation with $q=0.8$, total mass of $0.9
M_\odot$, showing equatorial slices
in the left column and slices perpendicular to this through the middle of
the core.
The top row shows logarithm
of density, the middle row shows temperature, and the bottom row shows the
$^{16}{\mathrm O}$  mass fraction. A blob is visible to the right of the merged core in the equatorial slices, and the vertical slices are taken through this blob.}
\label{brandon08hybrid}
\end{figure}

The SPH simulation is very different. There was no core-merger, and the
accretor core remains very much like the original accretor core, with the
donor material mostly smeared out around it. Contrary to the $q=0.5$ hybrid
SPH simulation, we do find that some $^{16}{\mathrm O}$ has been dredged up
in the $q=0.8$ hybrid SPH simulation.
We find that $5\times10^{-5} M_\odot$ of $^{16}{\mathrm
O}$ is at $\rho<10^{5.2}~{\rm g/cm^3}$ ($2\times10^{-5} M_\odot$ of
$^{16}{\mathrm O}$ at $\rho<10^5~{\rm g/cm^3}$). With an SPH particle
mass of $\sim4.5\times10^{-8}~{\rm M_\odot}$, this means that $\sim10^3$
CO SPH particles have been dredged up from the CO core of the hybrid
accretor.
The logarithm of density,
temperature, and $^{16}{\mathrm O}$ mass fraction is shown in
Fig.~\ref{brandon08hybrid}. A clear SOF formed around the core also in this
simulation, reaching temperatures above $2\times10^8~{\rm K}$, much higher
than the highest temperatures found in the grid-based simulations. 
As in the grid-based simulations, the equatorial plane is noticeably cooler 
than the rest of the SOF. The highest density found in this SPH simulation is 
$2\times10^6~{\rm g~cm^{-3}}$. 

Of interest is also that a blob does form, and is clearly visible in
Fig.~\ref{brandon08hybrid} to the right of the merged core in that figure.
It has the same features as in the grid-based simulations, that it is colder
and has higher density than the surroundings. Because so little
$^{16}{\mathrm O}$ has been dredged up, no clear feature can be seen in the
$^{16}{\mathrm O}$ mass fraction plots.
This blob persists to the end of the simulation, but it gets gradually
smeared out over time, as in the grid-based simulations.

\section{Discussion}
\label{discussionsection}

In this paper we are interested in estimating the amount of dredged-up CO
material from the accretor in mergers between a normal CO WD or a hybrid
CO/He WD and a He WD.  In \citet{staff12} we presented five hydrodynamics
simulations of the merger of a CO WD with a He WD with a range of mass
ratios between $q=0.5$ and $q=0.99$ and a total mass of $M_{\rm tot}=0.9
M_\odot$.  We found that in all cases much $^{16}\mathrm{O}$ was dredged up,
making it difficult to produce the observed oxygen ratio of order unity in
RCB stars during the dynamic merger phase.  In that paper, we also
speculated that the amount of dredged-up $^{16}\mathrm{O}$ could be limited
if the accretor is a ``hybrid'' CO/He WD \citep{rappaport09,iben85}, ie.  a
WD that has a CO core with a thick helium layer on top of it.  We have performed two
simulations with such a hybrid CO/He WD accretor, one with a $0.48 M_\odot$
accretor and a $0.24 M_\odot$ He WD donor (ie.  a mass ratio $q=0.5$), and
one with accretor mass of $0.5~{\rm M_\odot}$ and donor mass of $0.4~{\rm
M_\odot}$ (mass ratio of 0.8).
We find that in both of these simulations, thanks to the
thick outer layer of He on the accretor, less  $^{16}\mathrm{O}$ is being
dredged up than in simulations with a pure CO WD accretor.
In the $q=0.5$ SPH simulation we even find that absolutely no $^{16}\mathrm{O}$ is found at lower densities following the merger.

\begin{figure}   
\includegraphics[width=0.45\textwidth]{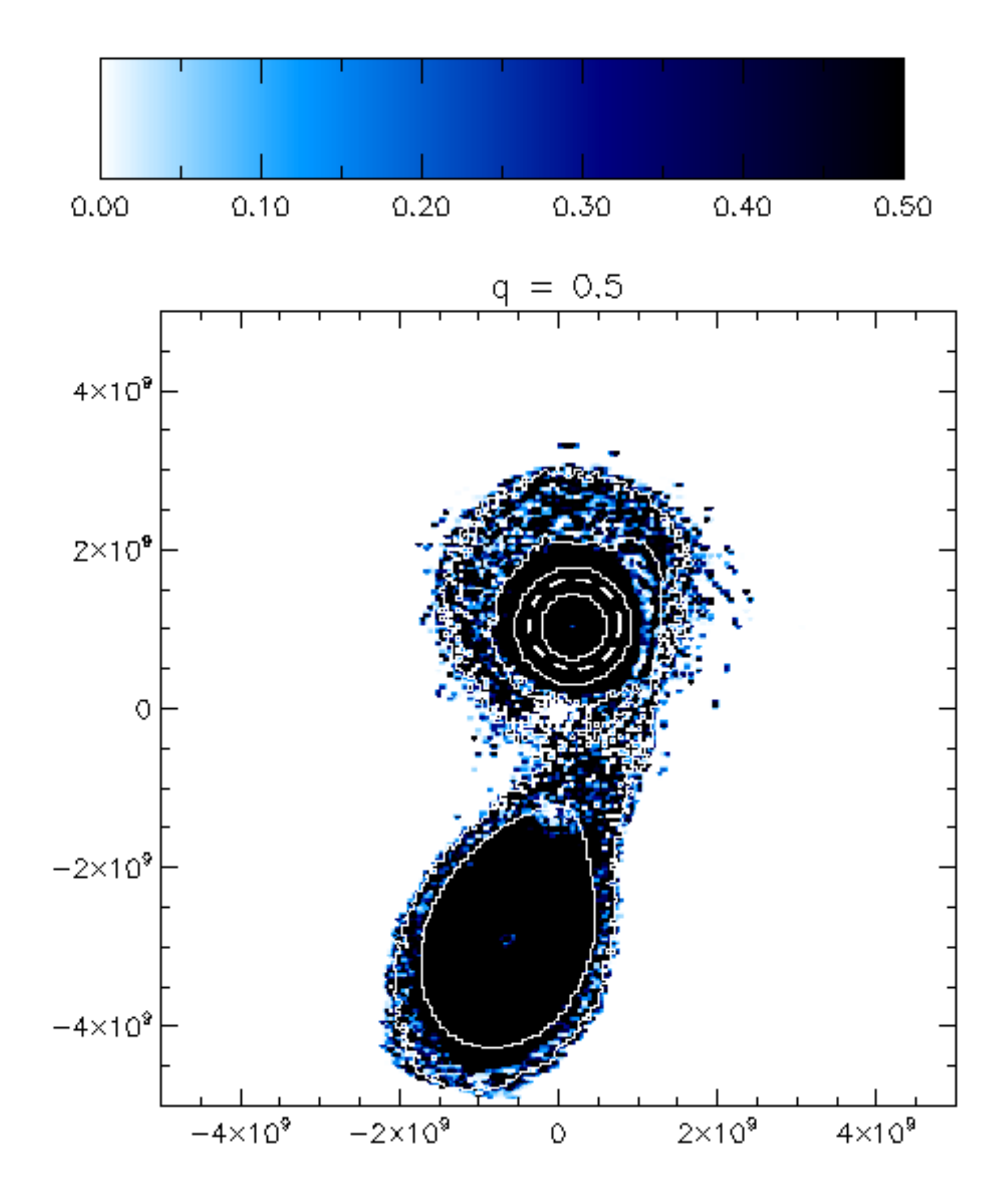}
\includegraphics[width=0.45\textwidth]{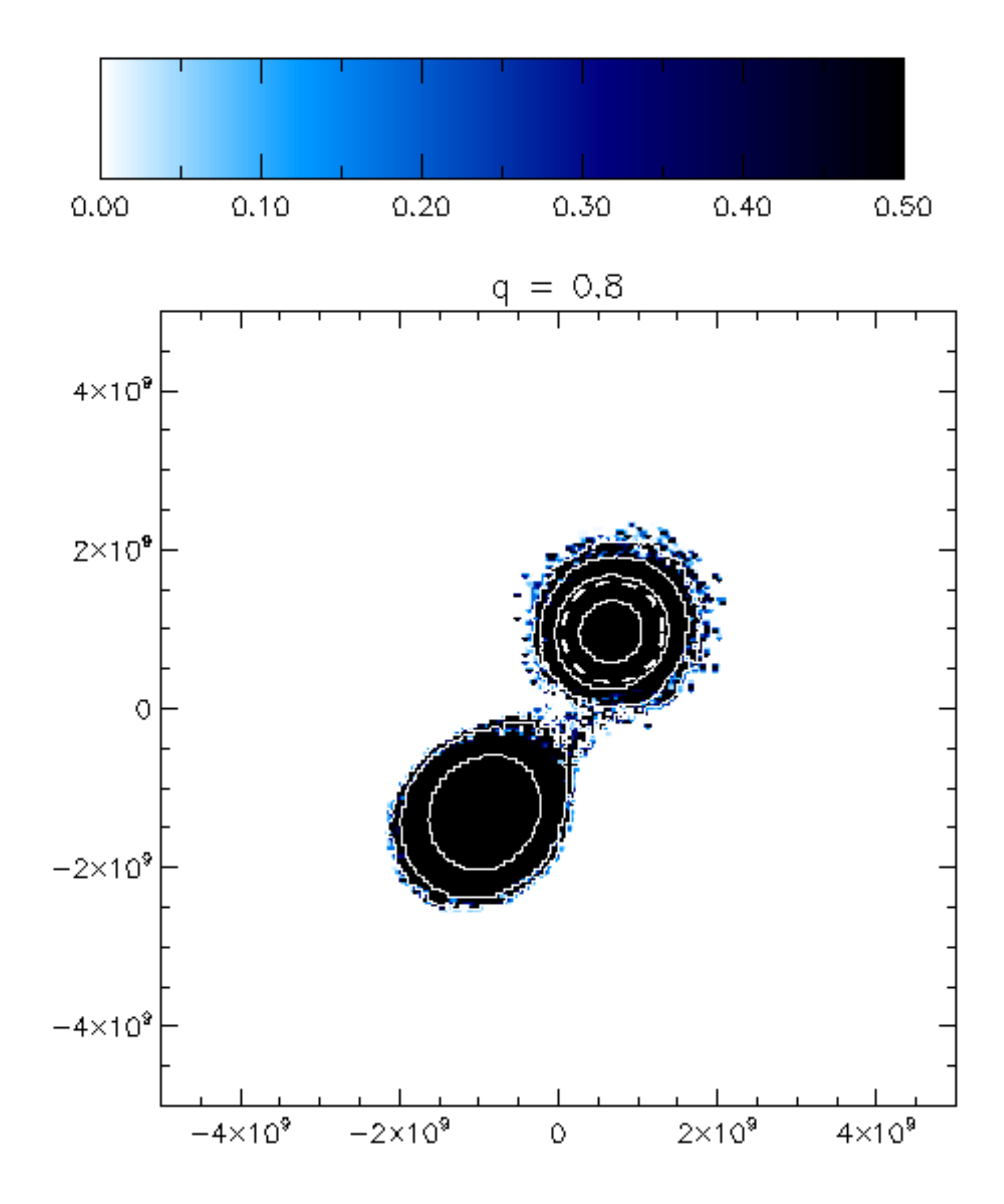}
\caption{Richardson number (color) and density contours in the equatorial plane for the $q=0.5$ SPH simulation (left panel) and the $q=0.8$ SPH simulation (right panel).
Black color indicates $R_i \geq 0.5$.
The snapshots are taken shortly before the final disruption of the donor.
The dashed line indicates the boundary of the CO core in the hybrid accretor, which has a radius of $5.7\times10^8~{\rm cm}$ in the $q=0.5$ simulation and a radius of $6.3\times10^8~{\rm cm}$ in the $q=0.8$ simulation.
The radius of the CO core remains reasonably constant in both simulations.
The axes are in centimeters.}
\label{richardsonnumber}
\end{figure}

We have calculated the Richardson number (see the appendix) in the hybrid SPH simulations, and found that indeed the low Richardson numbers are only in the helium layer (see Fig.~\ref{richardsonnumber}).
A low Richardson numbers indicates that there is sufficient free kinetic energy in the velocity sheer to overturn the fluid.
Hence, only helium is being dredged up from the hybrid accretor, to mix in with helium from the donor.
Though in the $q=0.8$ simulation, we see that there are low Richardson numbers near the CO core boundary around where the stream impacts the accretor, and it is not unthinkable that at another time these high Richardson numbers could be found slightly deeper, explaining how a small amount of $^{16}\mathrm{O}$ is eventually found at low densities in this simulation.

However, it turns out that in the grid-based hybrid simulations, even before
the merger, much CO material was found outside the CO core of the
accretor despite the overlying helium layer.  
Within roughly one orbit, several percent of a solar mass of accretor material is found at densities below the threshold densities.
It may be that there is artificial (numerical) diffusion of CO accretor material out into the SOF, and therefore we decided to reset the mass fractions to artificially ensure that all the CO material was in the core (as discussed in section 2.1).  
Even with this
numerical trick, we find that much CO material is still being brought to lower densities during the merger.  
With the hybrid accretor and $q=0.5$ in the grid-based simulations, we find that $0.01-0.03~{\rm M_\odot}$ of $^{16}{\mathrm O}$ finds its way out of the high density core and to the SOF.

We did not do this resetting of mass fractions in the $q=0.8$ simulation, in
part explaining why we find more $^{16}{\mathrm O}$ at lower densities in
that simulation.  However, due to the violent core merger occurring in this
case\footnote{By a ``less violent merger'', we mean a merger where the donor
material ``gently'' ends up on top of the accretor, leaving the accretor
relatively unchanged except from the outermost part.  A ``violent merger''
then, is one where the accretor core is changed significantly.}, it is not
surprising that we find that much material from the core is being brought
out to lower densities.
The $q=0.8$ SPH simulation did not result in such a violent core merger, and instead the merger behaved more like the lower q simulations, where the donor material more gently puts itself on top of the accretor.

Furthermore, secondary effects (for instance squeezing of the accretor due
to the impact of the accretion stream) also do not lead to dredge up of CO
material from deep in the core of the accretor in the SPH simulations. 
The SPH simulations do not suffer from the artificial diffusion of the mass
fractions that we suspect may cause the large dredge-up effect in the grid
based codes, since the mass fractions are inherent to each SPH particle. 
The SPH
simulations may, however, underestimate the amount of dredge up, since in
these simulations the smallest mass unit that can be dredged up is one SPH
particle. One SPH particle is slightly less than $10^{26}~{\rm g}$, depending
on the simulation. We do emphasize that dredge up is possible also in the
SPH simulations, as observed in the $q=0.7$ and $q=0.8$ SPH simulations.

An important difference between SPH and grid-based simulations is the
treatment of viscosity in the two types of codes, and this may well affect
the amount of mass being dredged up. The SPH code only applies viscosity in
converging flows, while the grid-based codes applies viscosity over all
discontinuities and local extrema. This would mean, for example, that the
SPH code does not apply viscosity at contact discontinuities, while the grid
based codes do.

Another aspect that affects the amount of $^{16}{\mathrm O}$ being
dredged up is our assumption that the CO WD contains 50\% C and 50\% O,
uniformly distributed throughout the star (or the CO core of the hybrid stars). In fact, CO WDs likely contains
more $^{16}{\mathrm O}$ than $^{12}{\mathrm C}$, but it will not be
uniformly distributed in the star. 
The internal $^{16}{\mathrm O}$ profile depends on the rate of the $^{12}{\mathrm C}(\alpha,\gamma)^{16}{\mathrm O}$ reaction \citep{salaris97}.
This leads to oxygen rich cores surrounded by a carbon rich layer
\citep[see a model of a $0.58~{\rm M_\odot}$ CO
WD in][]{staff12}.
By using that model, instead of assuming 50\% C and 50\% O uniformly distributed, we found that the amount of $^{16}{\mathrm O}$ in the SOF was reduced a factor of $\sim2$ in the $q=0.7$ simulation (the difference was smaller in the other models that we used).
Hence the amount of $^{16}{\mathrm O}$ that we find (which is listed in Table~\ref{jansimrestable}) might be overestimated by up to a factor of $\sim2$.

The resulting density distribution looks similar between the non-AMR and AMR grid-based
simulations, but the grid-based simulations do not show a separation of donor and
accretor material in the same way as we found in the $q=0.7$ SPH simulation.  
In all the grid simulations (see
e.g. Fig.~\ref{jan07}) we find the blob \citep[discussed in][]{staff12}, 
which is donor rich.  In some of the SPH
simulations a blob-like structure is visible for a short time after the
merger, but this disappears over time and the core appears quite
axisymmetric at the end of the simulation.  This feature is visible around
``3-4 o'clock'' from the merged core in the equatorial slice in
Fig.~\ref{brandon07}, as a colder and slightly denser structure. It is also
visible in the vertical density plot, as the $10^4-10^5~{\rm g~cm^{-3}}$
contour level is elongated to the right. Since very
little accretor material ends up in the equatorial plane in this SPH
simulation, this blob-like structure is not associated with a gas that is
poorer in accretor material than the surroundings, as we see in the
grid-based simulations. Whether this is the reason for the disappearance of
the blob in the SPH simulations remains unclear, but it is supported by the
fact that we find no such blob in the $q=0.5$ SPH simulation, where we also
find no dredge up of $^{16}{\mathrm O}$ from the core.

We find that the maximum density 
in the core of the merged object slightly higher in the non-AMR simulations
than in the AMR simulations, and even higher in the SPH simulations. Even initially, the maximum density is slightly higher in the SPH simulations. This is likely resolution dependent, since SPH simulations have higher resolution at higher density, and therefore can resolve the core better. The
temperature in the SOF is also higher in the non-AMR simulations, and even higher in
the SPH simulations. In the SOF,
the equatorial plane is noticeably cooler than the rest of the SOF, a
feature seen in all the simulations forming a SOF.
Another noticeable
difference is that the core in the AMR simulations appears to remain more
pure CO compared with the non-AMR simulation. This is likely due to the
diffusion of the mass fractions mentioned in \citet{staff12}, which may be
limited with the higher resolution in the AMR simulations. The merged core
in the SPH simulations remains very CO pure.

We have talked about much $^{16}\mathrm{O}$ being dredged up,
and one may reasonably wonder how much is too much dredge-up of 
$^{16}\mathrm{O}$ in the context of RCB stars?  Based on \citet{zhang14} we
can find the amount of $^{18}\mathrm{O}$ that may be synthesized. Since the
oxygen ratio should be of order unity, no more than that amount of
$^{16}\mathrm{O}$ should be dredged up, as the $^{16}\mathrm{O}$ is not
being destroyed \citep{staff12}, and this assumes that no significant
amount of $^{16}\mathrm{O}$ is being produced. 
\citet{zhang14} found that in most of their models the surface oxygen mass
fraction is 0.005 to 0.008 (see their tables 2 and 3).  In most of their
models they found the $^{16}\mathrm{O}$ to $^{18}\mathrm{O}$ ratio to be of
order unity (comparable to observations), so roughly half of this is
$^{16}\mathrm{O}$.  The envelope mass is roughly equivalent to the mass of
the He WD donor, or $\sim0.3~{\rm M_\odot}$.  Hence the $^{16}\mathrm{O}$
and $^{18}\mathrm{O}$ mass in the envelope is $\sim10^{-3}~{\rm M_\odot}$,
and this is therefore roughly how much $^{18}\mathrm{O}$ \citet{zhang14}
found could be synthesized.  In our grid-based simulations we find more than
ten times as much $^{16}\mathrm{O}$ being dredged up in the $q=0.5$
simulation (the ``best case''), and even more in the higher q simulations
(see Table~\ref{jansimrestable}). 
In the hybrid SPH simulations, however, no or very little $^{16}\mathrm{O}$ is dredged up, and these could therefore be excellent candidates for producing the oxygen ratio seen in RCB stars.

\citet{dan14} ran 225 SPH
simulations of WD mergers with varying masses, total masses, mass ratios,
composition, etc.
Relevant to this paper they find that not much mass is lost in the
merger event, up to only $3.4\times10^{-2}M_\odot$, where increasing mass
ratio leads to less mass ejected.  They also investigate the degree of
mixing of donor and accretor material, and find that for $q\lesssim0.45$
there is hardly any mixing, while most mixing occurs for near equal mass
ratios.  We have not done $q<0.5$, but this result is qualitatively in
agreement with what we found in \citet{staff12}, that for $q\sim1$ the
merged core consist of a mix of donor and accretor material, while for
$q=0.5$ the core after merger consists mainly of accretor material.
\citet{dan14}
also find that for $M_{\rm tot}\lesssim1M_\odot$ (which is the mass range
that we focus on), nuclear burning is negligible on the time scale that they
simulate, in agreement with our SPH simulations.  This is 
likely because they, like us, have not included hydrogen in their
simulations\footnote{The He WDs have a low mass layer of hydrogen on top of
the WD.}, as we found in \citet{staff12} that it can react very rapidly,
releasing much energy.  In that paper, we also found that while helium
burning can occur, it is on a timescale that is long compared to the
dynamical timescale that we simulate.  

\citet{zhu13} performed a parameter study of the
merger of unsynchronized CO WDs. They found many results similar to 
\citet{dan14}, although the fact that the stars in \citet{zhu13} are not
tidally locked results in differences regarding the amount of mixing
compared with the simulations in \citet{dan14} who, like us, 
simulates WDs that are initially synchronously rotating.

The nucleosynthesis that can form $^{18}{\mathrm O}$ will occur in the SOF (or possibly in hot-spots in the core; see below),
which is located just outside the core at a radius of roughly $10^9~{\rm
cm}$.  If these mergers are to produce RCB stars, they will have to swell up
to giant sizes ($\sim10^{13}~{\rm cm}$).  The $^{18}{\mathrm O}$ and other
elements produced in the SOF will then have to be brought up to the
surface.  As $^{18}{\mathrm O}$ is produced in the SOF, it may become
buoyant and therefore be transported out of the SOF to regions of the star
where nucleosynthesis cannot occur.  If many elements have already been
dredged-up to areas outside of the SOF, this can be mixed in with the newly
formed elements in the SOF.  Of relevance for the oxygen ratio, it is
therefore not only the oxygen ratio in the SOF that is important, as the
oxygen ratio might be further diluted if much $^{16}{\mathrm O}$ is present
outside the SOF.
The strength of the mixing and depth at which elements are brought up from (i.e. from the core, from the SOF,...) following the merger is therefore of great relevance. 
\citet{menon13} found that both the magnitude of the mixing and the depth must decrease over time in order to get the observed elemental abundances found in RCBs.

The depth of the mixing relates to the question of how to define the core in the simulations. 
In this work we have implemented two density thresholds ($\rho=10^5~{\rm g~cm^{-3}}$ and $\rho=10^{5.2}~{\rm g~cm^{-3}}$) and found the amount of $^{16}{\mathrm O}$ below these density thresholds.
We have then assumed that all this $^{16}{\mathrm O}$ is being brought into the envelope and affects the oxygen ratio, as the $^{18}{\mathrm O}$ would also have to be produced in the same region.
While the nucleosynthesis processes will likely proceed faster in hotter and denser environments (in the SOF), that does not necessarily result in more $^{18}{\mathrm O}$ being formed there, as it can also be destroyed.
$^{16}{\mathrm O}$ is also produced in this region, but as we found in \citet{staff12} it is for the most part not being destroyed.
A more detailed study will need to be performed in order to understand exactly where in the SOF the $^{18}{\mathrm O}$ forms and how deep the mixing can go in order to obtain the observed oxygen ratios.
In this paper we focused on investigating how much $^{16}{\mathrm O}$ is being brought up during the dynamical merger to the region where it can affect the observed oxygen ratio, and we chose these density thresholds to get some feel for the importance of the depth.
We note that these density thresholds for the core are not inconsistent 
with the core boundaries found in \citet{paxton13} for a somewhat more 
massive WD than what we use here.

In \citet{staff12} we did not consider the high-q simulations when
discussing possible nucleosynthesis and the resulting element abundances, as
these simulations did not form a SOF. However, hot spots may still develop
in the cores of these simulations as in the grid-based $q=0.8$ simulation. 
\citet{clayton07} discussed the importance of a small amount of hydrogen to the nucleosynthesis. Can the
resulting energy release further heat up these hot spots, allowing for
helium to react? In these high-q simulations, the cores merged causing much
dredge-up of $^{16}{\mathrm O}$, and therefore it seems difficult to form
sufficient $^{18}{\mathrm O}$ to achieve the oxygen ratios required for RCB
stars. What will such high-q merged object then look like? In order to
begin answering this question, properly resolving the hydrogen in the outer
part of the He WD will be necessary, as well as treating the nuclear
reactions, including hydrogen, in the hydrodynamics simulations.

In the simulations presented in this paper, we have ignored the effects of
magnetic fields.  Magnetic fields, however, may suppress Kelvin-Helmholz
instabilities from growing (for instance on the surface between the SOF and
the core), thereby possibly reducing the amount of dredge-up of
$^{16}{\mathrm O}$ from the core.  This may make it easier to achieve the
oxygen ratios.

\section{Summary}
\label{conclusionsection}

\begin{table}
\caption{Summary of the simulations showing the mass ratio and type, and 
the main results showing amount of $^{16}{\mathrm O}$ dredged up to lower
densities, the maximum temperature found in the SOF during and after the
merger (or in hot spots in the merged core in the $q=0.8$ grid-based
simulations), the mean temperature in the SOF (or hot spots), and estimates of the mean density in the SOF (or hot spot).}
\begin{tabular}{ccccccc}
\hline
simulation & total mass & $^{16}{\mathrm O}$ at & 
$^{16}{\mathrm O}$ at & max(T) in & $\overline{T}$ & $\overline{\rho}$
\\
type and & & $\rho<10^{5.2}~{\rm g~cm^{-3}}$  & $\rho<10^{5}~{\rm g~cm^{-3}}$
& SOF & SOF & SOF\\
mass ratio & $[M_\odot]$ & $[M_\odot]$ &  $[M_\odot]$ & $[{\rm K}]$ & $[{\rm K}]$ & $[{\rm g~cm^{-3}}]$  \\
\hline
$q=0.7$:\\
non AMR & $0.9$ & $0.045$ & $0.035$ & $1.5\times10^8$ & $1.1\times10^8$ & $4.3\times10^4$ \\
AMR & $0.9$ & $0.04$ & $0.03$ & $7.0\times10^7$ & $9.3\times10^7$ & $3.3\times10^5$ \\ 
SPH  & $0.9$ & $0.02$ & $0.01$ & $2.5\times10^8$ & $1.9\times10^8$ & $1.1\times10^5$ \\
\hline
$q=0.5$:\\
non AMR  & $0.71$ & $0.02$ & $0.012$ & $2.1\times10^8$ & $1.6\times10^8$ & $2.3\times10^4$\\
AMR  & $0.758$ &  $<0.035$ & $<0.03$ & $1.5\times10^8$ & $1.2\times10^8$ & $5.5\times10^4$ \\ 
SPH  & $0.71$ & $0.0$ & $0.0$ & $2.5\times10^8$ & $2.0\times10^8$ & $4.6\times10^4$ \\ 
\hline
$q=0.8$:\\
non AMR  & $0.9$ & $0.044$ & $0.03$ & $1.25\times10^8$ & $1.1\times10^8$ & $8.5\times10^4$ \\
(hot spot) \\
AMR  & $1.01$ & $0.06$ & $0.04$ & $8.1\times10^7$ & $8.0\times10^7$ & $6.4\times10^5$ \\ 
(hot spot)\\
SPH (SOF) & $0.9$ & $5\times10^{-5}$ & $2\times10^{-5}$ & $2.0\times10^8$ & $1.8\times10^8$ & $1.0\times10^5$ \\
\hline
\end{tabular}
\label{jansimrestable}
\end{table}

We have run three different WD merger simulations, using three different
codes: a non-AMR grid-based hydrodynamics code on a cylindrical grid, an 
AMR grid-based hydrodynamics code on a Cartesian grid, and a smooth 
particle hydrodynamics code.  We summarize
the simulations and the main results in Table~\ref{jansimrestable}.  Between
the two grid-based codes we find very good agreement, and they also agree reasonably
well with the SPH code.  However, in the SPH simulations we find much less $^{16}{\mathrm O}$ (accretor material) 
at lower densities than in the grid-based codes.  Also, we found that in the $q=0.8$
simulation, the two cores merged in the grid-based simulations and no SOF
was formed. Instead, ``hot spots'' formed in and around the merged core. 
In the SPH simulation, we find no He in the merged core
indicating that the cores did not merge, and some kind of a SOF does form 
although it is much cooler in the equatorial plane than elsewhere.

The temperature in the $q=0.8$ simulations does not get very
high in either of the grid-based simulations in high density regions, and little 
nucleosynthesis can occur. In the SPH simulation, however, temperatures of $\sim2\times10^8~{\rm K}$ is found in an SOF above and below the equatorial plane, making nucleosynthesis possible.  The $q=0.5$ simulation with a
hybrid accretor may be the most interesting in the context of formation of
RCB stars, as temperatures of the order of $1.5-2\times10^8~{\rm K}$ is found in
the SOF, at densities of the order $10^5~{\rm g~cm^{-3}}$.  This is
sufficiently hot and dense that nucleosynthesis processes including helium
burning can occur.  We also found in the SPH simulation that no accretor
material is dredged up to lower densities, allowing for oxygen ratios of
order unity if equal amounts of $^{18}{\mathrm O}$ is produced in the SOF.
Not much nucleosynthesis is expected in the $q=0.8$ SPH simulation, and 
therefore it is likely not favorable for producing the high oxygen ratio even
though not much $^{16}{\mathrm O}$ is being dredged up. However, if the SOF 
becomes hotter with time in the high density regions, nucleosynthesis may 
occur, which could lead to the production of $^{18}{\mathrm O}$ and 
consequently a high oxygen ratio. 

In the grid-based simulations, we find that much $^{16}{\mathrm O}$ is being
dredged up from the accretor during the merger event, even if the accretor
is a hybrid CO/He WD with a thick $>0.1~{\rm M_\odot}$ layer of 
He on top.  
In fact, we found that right from the begining of the simulation the amount of $^{16}{\mathrm O}$ at lower densities grows rapidly, indicating that this is a numerical effect.
In the non-AMR $q=0.5$ simulation, we tried
to artificially ``reset'' the hybrid accretor shortly before the merger
event to ensure that no accretor material was at lower densities then.
Despite this, we still found much $^{16}{\mathrm O}$ at densities where
nucleosynthesis could occur, making it very difficult to reach the observed
oxygen ratios.

It seems clear that the grid-based codes overestimate the amount of dredge-up,
due to the artificial diffusion of the mass fractions. 
The SPH code, run in very high resolution ($\sim20\times10^6$ particles) may 
more accurately track the amount of dredge-up, and our conclusion is that 
a hybrid accretor with a thick outer layer of helium can prevent dredge-up of 
oxygen from the core. A merger between a He WD and a hybrid CO/He WD with a
mass ratio ($q\lesssim0.8$; in which the cores do not merge and an SOF forms) is
therefore a good candidate for the progenitor system of RCB stars. This would, however, indicate that the mass of RCB stars should be significantly lower than $1~{\rm M_\odot}$, since the hybrid CO/He WDs have a maximum mass of $<0.5~{\rm M_\odot}$.

\acknowledgements
We would like to thank J. E. Tohline for helpful discussions, and the anonymous referee for many helpful comments.
Some of the work was done by J.E.S. while he was a post doc. at
Macquarie University and at the University of Florida, and he
acknowledges support from the Australian Research Council Discovery 
Project (DP12013337) program. 
This work has been supported, in part, by grant NNX10AC72G from NASA's ATP 
program, and in part by NASA grant NNX15AP95A.
We wish to acknowledge the support from the National Science Foundation
through CREATIV grant AST-1240655.
The computations were carried out in part using the computational resources
of the Louisiana Optical Network Initiative (LONI), and the XSEDE 
machine Kraken through grant TG-AST090104 and TG-AST110034.
LANL simulations were carried out on Turquoise network platform Wolf under Institutional Computing (IC) allocations. One simulation was carried out on UVI's Bucc cluster.

\appendix

\section{Measurement of the Richardson Number}

The Richardson number \citep[$Ri$;][]{drazin04} is given by:
\begin{equation*}
   Ri = - \frac{ \nabla \Phi_{eff} \nabla \rho}{\rho \left( \nabla v \right)^{2} }.
\end{equation*}

Where $Ri<1/4$, the flow can over-turn so these are regions where the Kelvin-Helmholtz instability can mix the fluid.
We have calculated the Richardson number in the SPH simulations. The SPH data had been put on a grid for this calculation.

The code measures the center of mass of each component, the accretor has center of mass coordinates in the
equatorial plane given by $x_{2}, y_{2}$. The velocity field is transformed to one where the accretor's center of 
mass is at rest and we then use an effective potential that corresponds to this angular frequency
\begin{equation*}
    \Phi_{eff} = \Phi - \frac{1}{2} \Omega^{2} \left( \vec{\mathbf{r}} - \vec{\mathbf{r}}_{2} \right)^{2}
\end{equation*}
The directions ``perpendicular'' and ``parallel'' to the flow locally are constructed from the gradient of the effective
potential as a normal direction from
\begin{equation*}
   \hat{\mathbf{n}} = \frac{ \vec{\mathbf{\nabla}} \Phi_{eff}}{ \left| \vec{\mathbf{\nabla}} \Phi_{eff} \right| }
\end{equation*}
and a tangent direction parallel to the equipotentials such that
\begin{equation*}
   \hat{\mathbf{t}} \cdot \hat{\mathbf{n}} = 0
\end{equation*}
and both $\hat{\mathbf{n}}$ and $\hat{\mathbf{t}}$ must lie in the equatorial plane due to symmetry about
the equatorial plane.

The cylindrical coordinate system centered on the origin of the grid will have unit vectors given by
$\hat{\mathbf{e}}_{r}$ and $\hat{\mathbf{e}}_{\phi}$ in the equatorial plane.
\begin{equation*}
   \hat{\mathbf{e}}_{r_{x}} = \cos{\phi}
\end{equation*}

\begin{equation*}
   \hat{\mathbf{e}}_{r_{y}} = \sin{\phi}
\end{equation*}

\begin{equation*}
   \hat{\mathbf{e}}_{\phi_{x}} = - \sin{\phi}
\end{equation*}

\begin{equation*}
   \hat{\mathbf{e}}_{\phi_{y}} = \cos{\phi}
\end{equation*}

The gradients that go in the calculation of the Richardson number are then computed from the gradient in the code's
coordinate system projected with the local directions normal and tangential to equipotential curves in the equatorial plane
as
\begin{equation*}
   \nabla \Phi_{eff} =  \hat{\mathbf{n}} \cdot \vec{\mathbf{\nabla}} \Phi_{eff} = 
                        \left( \hat{\mathbf{n}}_{x} \cdot \hat{\mathbf{e}}_{r_{x}} + \hat{\mathbf{n}}_{y} \cdot \hat{\mathbf{e}}_{r_{y}} \right) \frac{ \partial \Phi_{eff}}{ \partial r} +
                        \left( \hat{\mathbf{n}}_{x} \cdot \hat{\mathbf{e}}_{\phi_{x}} + \hat{\mathbf{n}}_{y} \cdot \hat{\mathbf{e}}_{\phi_{y}}  \right) \frac{ \partial \Phi_{eff}}{r \partial \phi}
\end{equation*}

\begin{equation*}
   \nabla \rho = \hat{\mathbf{n}} \cdot \vec{\mathbf{\nabla}} \rho = 
                       \left( \hat{\mathbf{n}}_{x} \cdot \hat{\mathbf{e}}_{r_{x}} + \hat{\mathbf{n}}_{y} \cdot \hat{\mathbf{e}}_{r_{y}} \right) \frac{ \partial \rho}{ \partial r} +
                        \left( \hat{\mathbf{n}}_{x} \cdot \hat{\mathbf{e}}_{\phi_{x}} + \hat{\mathbf{n}}_{y} \cdot \hat{\mathbf{e}}_{\phi_{y}} \right) \frac{ \partial \rho}{r \partial \phi}
\end{equation*}

\begin{eqnarray*} 
   \nabla v = \hat{\mathbf{t}} \cdot \left( \hat{\mathbf{n}} \cdot \vec{\mathbf{\nabla}} \vec{\mathbf{v}} \right) & = &
                    \left( \hat{\mathbf{e}}_{r_{x}} \cdot \hat{\mathbf{t}}_{x} + \hat{\mathbf{e}}_{r_{y}} \cdot \hat{\mathbf{t}}_{y} \right) 
                     \left( \left( \hat{\mathbf{n}}_{x} \cdot \hat{\mathbf{e}}_{r_{x}} + \hat{\mathbf{n}}_{y} \cdot \hat{\mathbf{e}}_{r_{y}} \right) \frac{\partial v_{r}}{\partial r} 
                     + \left(  \hat{\mathbf{n}}_{x} \cdot \hat{\mathbf{e}}_{\phi_{x}} +  \hat{\mathbf{n}}_{y} \cdot + \hat{\mathbf{e}}_{\phi_{y}} \right) \frac{\partial  v_{r}}{ r \partial \phi} \right)\\
               &+ & \left( \hat{\mathbf{e}}_{\phi_{x}} \cdot \hat{\mathbf{t}}_{x} + \hat{\mathbf{e}}_{\phi_{y}} \cdot \hat{\mathbf{t}}_{y} \right)
               \left( \left( \hat{\mathbf{n}}_{x} \cdot \hat{\mathbf{e}}_{r_{x}} + \hat{\mathbf{n}}_{y} \cdot \hat{\mathbf{e}}_{r_{y}} \right) \frac{\partial v_{\phi}}{\partial r}
                  + \left( \hat{\mathbf{n}}_{x} \cdot \hat{\mathbf{e}}_{\phi_{x}} + \hat{\mathbf{n}}_{y} \cdot \hat{\mathbf{e}}_{\phi_{y}} \right) \frac{\partial v_{\phi}}{r \partial \phi} \right).
\end{eqnarray*}

The derivatives in the previous expressions were computed numerically from 4th ordered centered finite differences.

\end{document}